\documentstyle[12pt]{article}
\title{Existence and stability of bushes of vibrational modes for octahedral mechanical
systems with Lennard-Jones potential}
\author{G.M.~Chechin, A.V.~Gnezdilov and M.Yu.~Zekhtser \\
Rostov State University, Rostov-on-Don, Russia}
\date{}

\begin{document}
\maketitle

\begin{abstract}
The special nonlinear dynamical regimes, ``bushes of normal
modes", can exist in the $N$-particle Hamiltonian systems with
discrete symmetry [Physica D 117 (1998) 43]. The dimension of the
bush can be essentially less than that of the whole mechanical
system. One-dimensional bushes represent the similar nonlinear
normal modes introduced by Rosenberg. A given bush can be excited
by imposing the appropriate initial conditions, and the energy of
the initial excitation turns out to be trapped in this bush.

In the present paper, we consider all possible vibrational bushes
in the simple octahedral mechanical system and discuss their
stability under assumption that the interactions between particles
are described by the Lennard-Jones potential.
\end{abstract}

\section{Introduction}
\label{Introduction}

A new concept, ``bushes of normal modes", was introduced for
nonlinear mechanical systems with discrete symmetry in~\cite{Dan1,
Dan2}. A given bush represents a certain superposition of the
modes associated with different irreducible representations
(irreps) of the symmetry group~$G$ of the mechanical system in
equilibrium. The coefficients of this superposition are
time-dependent functions for which the exact ordinary differential
equations can be obtained. In this sense, the bush can be
considered as a dynamical object whose dimensionality is generally
less than that of the original mechanical system. The following
propositions were justified in previous
papers~\cite{Dan1,Dan2,PhysD}:
\begin{enumerate}
\item A certain subgroup $G_D$ of the symmetry group $G$
corresponds to a given bush, and this bush can be excited by
imposing the appropriate initial conditions with the above
symmetry group $G_D \subset G$.
\item Each mode belonging to the bush possesses its own symmetry
group which is greater than or equal to the group $G_D$ of the
whole bush.
\item In spite of evolving mode amplitudes, the complete
collection of modes in the given bush is preserved in time and, in
this sense, the bush can be considered as a geometrical object.
\item The energy of the initial excitation is trapped in the bush,
i.e. it cannot spread to the modes which do not belong to the
bush, because of the symmetry restrictions.
\item As an indivisible nonlinear object, the bush exists because of
force interactions between the modes contained in it.
\item Taking into account the concrete type of interactions
between particles of the considered mechanical system can only
reduce the dimension of the given bush.
\item The extension of the bush can be realized as a result of the
loss of its stability which is accompanied by spontaneous breaking
of the bush symmetry (dynamical analog of phase transition).
\end{enumerate}

The special group-theoretical methods for finding bushes of modes
are discussed in~\cite{PhysD,IJNM}. The computer implementation of
these methods~\cite{Iso} (see also~\cite{Chechin}) allowed us to
find bushes of modes for wide classes of mechanical systems with
discrete symmetry. In particular, ``irreducible" bushes of
vibrational modes and symmetry determined similar non-linear
normal modes for all N-particle mechanical systems with the
symmetry of any of the 230 space groups were found in~\cite{IJNM}.
The bushes of vibrational modes of small dimensions were found and
classified into universality classes for all mechanical systems
with point groups of crystallographic symmetry in ~\cite{ENOC}.

Unlike discussions of existence of bushes of modes, a few papers
were devoted to the problem of their stability (as an example, we
can refer to the investigation of bushes of vibrational
modes for FPU-$\alpha$ chains in~\cite{FPU}). Nevertheless, the
problem of stability is one of the most important for bush theory.
Indeed, we must keep in mind that bushes of modes can be
considered as new physical objects only in the case where they are
stable in the {\it finite} domains of pertinent parameters of the
appropriate dynamical systems.

The problem of existence of bushes of modes can be studied with
the aid of the group-theoretical methods only, and appropriate
results do not depend on the concrete type of interactions between
particles of our mechanical system. In contrast, studying of bush
stability depends on these interactions essentially.

In the present paper, after the outline of the bush theory in Sec.2,
we consider bushes of vibrational modes for an octahedral mechanical
system of mass particles depicted in Fig.~\ref{fig.1},
with or without a particle in the center of the octahedron. All
bushes of modes are presented in Sec.3. The differential equations
describing their dynamics are discussed in Sec.4. The stability of
the bushes of vibrational modes is investigated in
Sec.5 for the case of interactions between particles described by the
Lennard-Jones potential.

\section{Bushes of vibrational modes}
\label{bushes}

A detailed consideration of the theory of bushes of modes with
some theorems about their structure and with the appropriate
group-theoretical methods can be found in~\cite{PhysD}. An
extensive description of the computational algorithm for finding
bushes in arbitrary crystal structures is presented
in~\cite{IJNM}. Let us now give an outline of the bush theory
with emphasis on the ideas important for our further discussion.

\subsection{Geometrical aspects}
\label{Geometrical aspects}

We consider  classical Hamiltonian systems of $N$ mass points
moving near a single equilibrium state which can be
characterized by a certain point or space symmetry group $G$. Let
the $3\times N$-dimensional vector,
\begin{equation}
\label{eq1} {\bf X}(t)=({\bf x}_1(t),{\bf x}_2(t),\ldots,{\bf
x}_N(t)),
\end{equation}
describe the displacements ${\bf x}_i(t)$ of all particles of our
mechanical system from their equilibrium positions. (Here we
denote by the three-dimensional vector ${\bf x}_i$ the
displacement of the $i$-th particle along the $X,Y$ and $Z$ axes).

The vector ${\bf X}(t)$ can be written as a superposition of all
basis vectors $ {\mbox{\boldmath$\varphi$}}_i^{(j)}$ of the
irreducible representations $\Gamma_j$ of the above mentioned
symmetry group $G$:
\begin{equation}
\label{eq2}{\bf
X}(t)=\sum_{j,i}\mu_i^{(j)}(t){\mbox{\boldmath$\varphi$}}_i^{(j)}.
\end{equation}
The coefficients $\mu_i^{(j)}(t)$ of this superposition depend on
time $t$, while the $3\times N$-dimensional time-independent
vectors ${\mbox{\boldmath$\varphi$}}_i^{(j)}$ determine the
specific patterns of displacements of all particles of our
mechanical system.

Note that the basis vectors ${\mbox{\boldmath$\varphi$}}_i^{(j)}$
are often called ``symmetry-adapted coordinates". In particular,
they can be normal coordinates. Hereafter, the term, ``mode",
means an arbitrary superposition of basis vectors corresponding to
a given irrep $\Gamma_j$. As a result of this definition, every
term $\mu_i^{(j)}(t){\mbox{\boldmath$\varphi$}}_i^{(j)}$ in the
right-hand side of Eq.(\ref{eq2}) is also a mode of the irrep
$\Gamma_j$. Sometimes, for brevity, we will refer to
$\mu_i^{(j)}(t)$ as a mode, but a reader must imagine that this
time-dependent coefficient is multiplied by the appropriate
$3\times N$-dimensional vector
${\mbox{\boldmath$\varphi$}}_i^{(j)}$ to give the mode in the
exact sense. We can also speak about {\it vibrational} modes
because only such type of symmetry-adapted (normal) modes are
considered in the present paper.

Every dynamical regime of the considered mechanical system can be
described by the appropriate time-dependent vector ${\bf X}(t)$
which determines a definite instantaneous  configuration of the
system. On the other hand, each instantaneous configuration
possesses a certain symmetry group $G_D$ (in particular, this
group may be trivial: $G_D=1$) which is a subgroup of the symmetry
group $G$ of the system in equilibrium ($G_D\subseteq G$).
Moreover, we can also ascribe a certain symmetry group to each
basis vector ${\mbox{\boldmath$\varphi$}}_i^{(j)}$ and to each
mode corresponding to a given irrep $\Gamma_j$ (remember that a
mode is a superposition of such vectors!), because the definite
instantaneous configurations correspond to them. The group $G_D$
contains all symmetry elements of group $G$ whose action does not
change this configuration.

Let us introduce, as it is usual in group theory, the operators
$\hat g$ associated with elements $g$ of group $G$ ($g\in G$)
which act on $3\times N$-dimensional vectors ${\bf
X}(t)$.\footnote{The symmetry elements $g\in G$ act on the vectors
of three-dimensional Euclidean space.} All elements $g\in G$ for
which
\begin{equation}
\label{eq3}\hat g{\bf X}(t)={\bf X}(t)
\end{equation}
form a certain subgroup $G_D\subseteq G$, and a complete set of
the above operators $\hat g$ ($\forall g\in G_D$) represents the
group $\widehat G_D$.

It can be shown that the symmetry group $G_D$ is preserved in time
in the sense that its elements {\it cannot disappear} during time
evolution. Actually, this property is a consequence of the
principle of determinism in classical mechanics.\footnote{The
phenomenon of spontaneous breaking of symmetry of a given
dynamical regime will be considered in the next section.} Thus,
the equation $\hat g{\bf X}(t)={\bf X}(t)$ $(g\in G_D)$, or,
formally,
\begin{equation}
\label{eq4} \widehat G_D{\bf X}(t)={\bf X}(t)
\end{equation}
is valid for every time $t$. As a consequence, we can classify the
different dynamical regimes in our nonlinear dynamical system,
described by the vectors ${\bf X}(t)$ from
Eqs.(\ref{eq1},\ref{eq2}), with the aid of {\it symmetry groups}
corresponding to them.

Using Eq.(\ref{eq4}), we can obtain the similar {\it invariance
conditions} for each individual irrep $\Gamma_j$ of the group $G$
(see details in~\cite{PhysD}):
\begin{equation}
\label{eq5} (\Gamma_j\downarrow
G_D){\mbox{\boldmath$\mu$}}_j={\mbox{\boldmath$\mu$}}_j.
\end{equation}
Here $(\Gamma_j\downarrow G_D)$ is a restriction of the irrep
$\Gamma_j$ to the subgroup $G_D$ of the group $G$, and
${\mbox{\boldmath$\mu$}}_j=(\mu_1^{(j)},\ldots,\mu_{n_j}^{(j)})$
is an {\it invariant vector} of $\Gamma_j$ ($n_j$ is the dimension
of this irrep).

To find all modes contributing to a given dynamical regime with
symmetry group $G_D$, i.e., for the vector ${\bf X}(t)$ from
Eq.(\ref{eq2}), we must solve linear algebraic
equations~(\ref{eq5}) for each irrep $\Gamma_j$ of the group $G$.
As a result of this procedure, the invariant vector
${\mbox{\boldmath$\mu$}}_j$ for some irreps $\Gamma_j$ can turn
out to be equal to zero. Such irreps do not contribute to the
considered dynamical regime. On the other hand, some nonzero
invariant vectors ${\mbox{\boldmath$\mu$}}_j$ for multidimensional
irreps may be of a very specific form because of definite
relations between their components (for example, certain
components can be equal to each other, or differ only by sign).
The contributions to ${\bf X}(t)$ from such irreps $\Gamma_j$
possess special forms. In Tables~\ref{2a},~\ref{2b}, we find, for
instance, dynamical regimes for which the contributions from the
irrep $\Gamma_{10}$ of the group $O_h$ are of the forms
$\mu_1^{10}(t)[{\mbox{\boldmath$\varphi$}}_1^{(10)}-{\mbox{\boldmath$\varphi$}}_2^{(10)}]$
and
$\mu_1^{10}(t)[{\mbox{\boldmath$\varphi$}}_1^{(10)}+{\mbox{\boldmath$\varphi$}}_2^{(10)}+{\mbox{\boldmath$\varphi$}}_3^{(10)}]$.

\begin{table}[htb]
\caption{Bushes of vibrational modes and their stability domains
for the octahedral structure with Lennard-Jones potential}
\label{2a} $$
\begin{array}{|c|c|c|c|c|c|}
\hline &&&&& \\

 \mbox{N}& \mbox{Point}& \mbox{Generators}& \mbox{Dim.}&
 \mbox{Bush of modes}& \mbox{Range of bush}\\ &\mbox{group}&&&& \mbox{stability} \\[4pt]
 \hline

 1 & O_h & 13, 14,33 & 1 &\underline {\mu_1^{(1)}{\mbox{\boldmath$\varphi$}}_1^{(1)}} & R=0.001 \\[2pt]
 \hline

 2 & D_{4h} & 2, 14, 25 & 2 & \mu_1^{(1)}{\mbox{\boldmath$\varphi$}}_1^{(1)}+\underline {\mu_1^{(5)}{\mbox{\boldmath$\varphi$}}_1^{(5)}} & R=0.009 \\[2pt]
 \hline

  3 & D'_{2d} & 13, 38 & 3 & \mu_1^{(1)}{\mbox{\boldmath$\varphi$}}_1^{(1)}+\mu_1^{(5)}{\mbox{\boldmath$\varphi$}}_1^{(5)}+\underline{\mu_3^{(8)}{\mbox{\boldmath$\varphi$}}_3^{(8)}} & R=0.028 \\[2pt]
 \hline

 4 & C_{4v} & 14, 37 & 3 & \mu_1^{(1)}{\mbox{\boldmath$\varphi$}}_1^{(1)}+\mu_1^{(5)}{\mbox{\boldmath$\varphi$}}_1^{(5)}+\underline{\mu_3^{(10)}{\mbox{\boldmath$\varphi$}}_3^{(10)}} & R=0.003 \\[2pt]
 \hline

  5 & D'_{2h} & 4, 13, 25 & 3 & \mu_1^{(1)}{\mbox{\boldmath$\varphi$}}_1^{(1)}+\mu_1^{(5)}{\mbox{\boldmath$\varphi$}}_1^{(5)}+\underline{\mu_3^{(7)}{\mbox{\boldmath$\varphi$}}_3^{(7)}} & R=0.055 \\[2pt]
 \hline

  6 & C'_{2v} & 13, 28 & 5 &
  \mu_1^{(1)}{\mbox{\boldmath$\varphi$}}_1^{(1)}+\mu_1^{(5)}{\mbox{\boldmath$\varphi$}}_1^{(5)}+\mu_3^{(7)}{\mbox{\boldmath$\varphi$}}_3^{(7)}+& R_1=0.001 \\&&&&
  \underline {\mu_1^{(8)} \left( {\mbox{\boldmath$\varphi$}}_1^{(8)}+{\mbox{\boldmath$\varphi$}}_2^{(8)}
  \right)}
   +\underline {\mu_1^{(10)}\left({\mbox{\boldmath$\varphi$}}_1^{(10)}-{\mbox{\boldmath$\varphi$}}_2^{(10)}\right)}&R_2=0.002\\[6pt]
\hline

 7 & D_3 & 9, 13 & 3 &
  \mu_1^{(1)}{\mbox{\boldmath$\varphi$}}_1^{(1)}+
  \mu_1^{(7)} \left( {\mbox{\boldmath$\varphi$}}_1^{(7)}+{\mbox{\boldmath$\varphi$}}_2^{(7)}+{\mbox{\boldmath$\varphi$}}_3^{(7)}\right)
   +&\\ &&&&\underline
   {\mu_1^{(8)}\left({\mbox{\boldmath$\varphi$}}_1^{(8)}+{\mbox{\boldmath$\varphi$}}_2^{(8)}+{\mbox{\boldmath$\varphi$}}_3^{(8)}\right)}&R=0.002\\[6pt]
 \hline

  8 & C_{3v} & 9, 37 & 3 &
  \mu_1^{(1)}{\mbox{\boldmath$\varphi$}}_1^{(1)}+
  \mu_1^{(7)} \left( {\mbox{\boldmath$\varphi$}}_1^{(7)}+{\mbox{\boldmath$\varphi$}}_2^{(7)}+{\mbox{\boldmath$\varphi$}}_3^{(7)}\right)
   +&\\ &&&&\underline
   {\mu_1^{(10)}\left({\mbox{\boldmath$\varphi$}}_1^{(10)}+{\mbox{\boldmath$\varphi$}}_2^{(10)}+{\mbox{\boldmath$\varphi$}}_3^{(10)}\right)}&R=0.025\\[6pt]
 \hline

  9 & D_{3d} & 13, 33 & 2 &
  \mu_1^{(1)}{\mbox{\boldmath$\varphi$}}_1^{(1)}+
  \mu_1^{(7)}\underline
{\left(
{\mbox{\boldmath$\varphi$}}_1^{(7)}+{\mbox{\boldmath$\varphi$}}_2^{(7)}+{\mbox{\boldmath$\varphi$}}_3^{(7)}\right)}
   &R=0.078\\[6pt]
 \hline
\end{array}
$$
\end{table}

\begin{table}[htb]
\caption{Bushes of vibrational modes and their stability domains
for the octahedral structure with Lennard-Jones potential}
\label{2b} $$
\begin{array}{|c|c|c|c|c|c|}
\hline &&&&& \\

 \mbox{N}& \mbox{Point}& \mbox{Generators}& \mbox{Dim.}&
 \mbox{Bush of modes}& \mbox{Range of bush}\\ &\mbox{group}&&&& \mbox{stability} \\[4pt]
 \hline

10 & D_{2h} & 2, 4, 25 & 3 &
\mu_1^{(1)}{\mbox{\boldmath$\varphi$}}_1^{(1)}+\underline{\mu_1^{(5)}{\mbox{\boldmath$\varphi$}}_1^{(5)}+\mu_2^{(5)}{\mbox{\boldmath$\varphi$}}_2^{(5)}}
& R'=0.006 \\[2pt]
 \hline

 11 & D'_2 & 4, 13 & 4 &
\mu_1^{(1)}{\mbox{\boldmath$\varphi$}}_1^{(1)}+\mu_1^{(5)}{\mbox{\boldmath$\varphi$}}_1^{(5)}+\underline{\mu_3^{(7)}{\mbox{\boldmath$\varphi$}}_3^{(7)}+\mu_3^{(8)}{\mbox{\boldmath$\varphi$}}_3^{(8)}}
& R'=0.027 \\[2pt]
 \hline

 12 & C_{2v} & 4, 26 & 5 &
\mu_1^{(1)}{\mbox{\boldmath$\varphi$}}_1^{(1)}+\mu_1^{(5)}{\mbox{\boldmath$\varphi$}}_1^{(5)}+\mu_2^{(5)}{\mbox{\boldmath$\varphi$}}_2^{(5)}+&\\&&&&\underline{\mu_3^{(8)}{\mbox{\boldmath$\varphi$}}_3^{(8)}+\mu_3^{(10)}{\mbox{\boldmath$\varphi$}}_3^{(10)}}
& R'=0.001 \\[2pt]
 \hline

 13 & C''_{2v} & 4, 37 & 4 &
\mu_1^{(1)}{\mbox{\boldmath$\varphi$}}_1^{(1)}+\mu_1^{(5)}{\mbox{\boldmath$\varphi$}}_1^{(5)}+\underline{\mu_3^{(7)}{\mbox{\boldmath$\varphi$}}_3^{(7)}+\mu_3^{(10)}{\mbox{\boldmath$\varphi$}}_3^{(10)}}
& R'=0.003 \\[2pt]
 \hline

 14 & C'_{2} & 13 & 7 &
\mu_1^{(1)}{\mbox{\boldmath$\varphi$}}_1^{(1)}+\mu_1^{(5)}{\mbox{\boldmath$\varphi$}}_1^{(5)}+\mu_1^{(7)}\left({\mbox{\boldmath$\varphi$}}_1^{(7)}+{\mbox{\boldmath$\varphi$}}_2^{(7)}\right)+
&\\&&&&\mu_3^{(7)}{\mbox{\boldmath$\varphi$}}_3^{(7)}+\underline{\mu_1^{(8)}\left({\mbox{\boldmath$\varphi$}}_1^{(8)}+{\mbox{\boldmath$\varphi$}}_2^{(8)}\right)+}&\\&&&&\underline{\mu_3^{(8)}{\mbox{\boldmath$\varphi$}}_3^{(8)}}+\mu_1^{(10)}\left({\mbox{\boldmath$\varphi$}}_1^{(10)}-{\mbox{\boldmath$\varphi$}}_2^{(10)}\right)
& R'=0.001 \\[4pt]
 \hline

15 & C_{s} & 28 & 8 &
\mu_1^{(1)}{\mbox{\boldmath$\varphi$}}_1^{(1)}+\mu_1^{(5)}{\mbox{\boldmath$\varphi$}}_1^{(5)}+\mu_2^{(5)}{\mbox{\boldmath$\varphi$}}_2^{(5)}+\mu_3^{(7)}{\mbox{\boldmath$\varphi$}}_3^{(7)}+
&R'_1=0.013\\&&&&\underline{\mu_1^{(8)}{\mbox{\boldmath$\varphi$}}_1^{(8)}+\mu_2^{(8)}{\mbox{\boldmath$\varphi$}}_2^{(8)}}+\underline{\mu_1^{(10)}{\mbox{\boldmath$\varphi$}}_1^{(10)}+\mu_2^{(10)}{\mbox{\boldmath$\varphi$}}_2^{(10)}}
 & R'_2=0.002 \\[2pt]
 \hline

 16 & C'_{s} & 37 & 7 &
\mu_1^{(1)}{\mbox{\boldmath$\varphi$}}_1^{(1)}+\mu_1^{(5)}{\mbox{\boldmath$\varphi$}}_1^{(5)}+\mu_1^{(7)}\left({\mbox{\boldmath$\varphi$}}_1^{(7)}+{\mbox{\boldmath$\varphi$}}_2^{(7)}\right)+
&\\&&&&\mu_3^{(7)}{\mbox{\boldmath$\varphi$}}_3^{(7)}+\mu_1^{(8)}\left({\mbox{\boldmath$\varphi$}}_1^{(8)}-{\mbox{\boldmath$\varphi$}}_2^{(8)}\right)+&\\&&&&\underline{\mu_1^{(10)}\left({\mbox{\boldmath$\varphi$}}_1^{(10)}+{\mbox{\boldmath$\varphi$}}_2^{(10)}\right)+\mu_3^{(10)}{\mbox{\boldmath$\varphi$}}_3^{(10)}}
& R'=0.007 \\[6pt]
 \hline

 17 & C_{3} & 9 & 4 &
\mu_1^{(1)}{\mbox{\boldmath$\varphi$}}_1^{(1)}+\mu_1^{(7)}\left({\mbox{\boldmath$\varphi$}}_1^{(7)}+{\mbox{\boldmath$\varphi$}}_2^{(7)}+{\mbox{\boldmath$\varphi$}}_3^{(7)}\right)+
&\\&&&&\underline{\mu_1^{(8)}\left({\mbox{\boldmath$\varphi$}}_1^{(8)}+{\mbox{\boldmath$\varphi$}}_2^{(8)}+{\mbox{\boldmath$\varphi$}}_3^{(8)}\right)+}&\\&&&&\underline{
\mu_1^{(10)}\left({\mbox{\boldmath$\varphi$}}_1^{(10)}+{\mbox{\boldmath$\varphi$}}_2^{(10)}+{\mbox{\boldmath$\varphi$}}_3^{(10)}\right)}
& R'=0.005 \\[6pt]
 \hline

18 & C_{i} & 25 & 6 &
\mu_1^{(1)}{\mbox{\boldmath$\varphi$}}_1^{(1)}+\mu_1^{(5)}{\mbox{\boldmath$\varphi$}}_1^{(5)}+\mu_2^{(5)}{\mbox{\boldmath$\varphi$}}_2^{(5)}+
&\\&&&&\underline{\mu_1^{(7)}{\mbox{\boldmath$\varphi$}}_1^{(7)}+\mu_2^{(7)}{\mbox{\boldmath$\varphi$}}_2^{(7)}+\mu_3^{(7)}{\mbox{\boldmath$\varphi$}}_3^{(7)}}
 & R''=0.001 \\[2pt]
 \hline
\end{array}
$$
\end{table}

Actually, Eq.(\ref{eq5}) can be considered as a source of certain
selection rules for spreading excitation from the root mode to a
number of other (secondary) modes. Indeed, if a certain mode with
the symmetry group  $G_D$ is excited at the initial instant (we
call it the ``root" mode), this group determines the symmetry of
the whole bush. The condition that the appropriate dynamical
regime ${\bf X}(t)$ must be invariant under the action of the
above group $G_D$ leads to Eq.(\ref{eq4}) and then to
Eq.(\ref{eq5}). If the vector ${\mbox{\boldmath$\mu$}}_j$ for a
given irrep $\Gamma_j$ proves to be a zero vector, then there are
no modes belonging to this irrep which contribute to ${\bf X}(t)$,
i.e., the initial excitation {\it cannot spread} from the root
mode to the secondary modes associated with the irrep $\Gamma_j$.

Note that basis vectors associated with a given irrep $\Gamma_j$
in Eq.(\ref{eq2}) turn out to be equal to zero when this irrep is
not contained in the decomposition of the full vibrational irrep
$\Gamma$ into its irreducible parts $\Gamma_j$. This is a source
of the additional selection rules which reduce  the  number of
possible vibrational modes in the considered bush. Trying every
irrep $\Gamma_j$ in Eq.(\ref{eq5}) and analyzing the above
mentioned decomposition of the vibrational representation
$\Gamma$, we obtain the whole bush of modes with the symmetry
group $G_D$  in the explicit form.

Actually, we have just outlined the group-theoretical scheme for
finding bushes of vibrational modes for arbitrary mechanical
systems. The detailed consideration of this scheme can be found
in~\cite{PhysD,IJNM}.

In conclusion, let us rewrite the vector ${\bf X}(t)$ from
Eq.(\ref{eq2}) as the sum of contributions
${\mbox{\boldmath$\Delta$}}_j$ from all different irreps
$\Gamma_j$ of the group $G$. For each irrep $\Gamma_j$ with the
dimension $n_j$, we introduce a ``supervector"
\begin{equation}
\label{eq20_0}
{\mbox{\boldmath$\Phi$}}_j=\lbrace{\mbox{\boldmath$\varphi$}}_1^{(j)},\ldots,{\mbox{\boldmath$\varphi$}}_{n_j}^{(j)}\rbrace
\end{equation}
which represents the complete set of basis vectors
${\mbox{\boldmath$\varphi$}}_i^{(j)}\; (i=1,\ldots,n_j)$ of this
irrep. Remembering that the invariant vector
${\mbox{\boldmath$\mu$}}_j$ has the form,
\begin{equation}
\label{eq21}
{\mbox{\boldmath$\mu$}}_j=\lbrace\mu_1^{(j)},\ldots,\mu_{n_j}^{(j)}\rbrace,
\end{equation}
we can rewrite Eq.(\ref{eq2}) in the following compact form:
\begin{equation}
\label{eq22} {\bf X}(t)=\sum_j {\mbox{\boldmath$\Delta$}}_j=\sum_j
\left(
{\mbox{\boldmath$\mu$}}_j(t)|{\mbox{\boldmath$\Phi$}}_j\right).
\end{equation}
Here we introduced a formal ``scalar" product $$\left(
{\mbox{\boldmath$\mu$}}_j(t)|{\mbox{\boldmath$\Phi$}}_j\right)=
\sum_{i=1}^{n_j}\mu_i^{(j)}(t){\mbox{\boldmath$\varphi$}}_i^{(j)}$$
of $n_j$-dimensional vectors ${\mbox{\boldmath$\mu$}}_j$ and
${\mbox{\boldmath$\Phi$}}_j$ (actually, this product represents a
certain $3\times N$-dimensional vector
${\mbox{\boldmath$\Delta$}}_j$!).

\subsection{Dynamical aspects}
\label{Dynamical aspects}

Let us return to Eq.(\ref{eq2}). We speak about geometrical
aspects of the bush theory when concentrating our attention on basis
vectors ${\mbox{\boldmath$\varphi$}}_i^{(j)}$, and we speak about
dynamical aspects of this theory when we focus on time-dependent
coefficients $\mu_i^{(j)}(t)$, which will be also called ``modes"
(see comments on this term in Sec.\ref{Geometrical aspects}).

If interactions between the particles of our mechanical system are
known, {\it exact} dynamical equations describing the time
evolution of a given bush can be written. We will consider this
question in Sec.\ref{exact} of the present paper, while some
general statements about the theory of bush dynamics are considered
here.

Two types of interactions between modes in nonlinear Hamiltonian
system are discussed in~\cite{PhysD}, namely, {\it force}
interactions and {\it parametric} interactions. We can illustrate
the difference between these types of modal interactions using a
simple example.

Let us consider two different linear oscillators whose coupling is
described by only one anharmonic term, $U=-\gamma \mu_1^2\mu_2$,
in the potential energy. Dynamical equations for this system can
be written as follows:
\begin{eqnarray}
\label{eq6a} \ddot\mu_1+\omega_1^2\mu_1&=&2\gamma\mu_1\mu_2, \\
\label{eq6b} \ddot\mu_2+\omega_2^2\mu_2&=&\gamma\mu_1^2.
\end{eqnarray}
Here $\gamma$ is an arbitrary constant characterizing the strength
of the interaction of the oscillators. We can suppose that
Eqs.(\ref{eq6a},\ref{eq6b}) describe the dynamics of two modes
$\mu_1(t)$ and $\mu_2(t)$ in a certain mechanical system.

An essential disparity between modes $\mu_1(t)$ and $\mu_2(t)$ can
be seen from the above equations. Indeed, if we excite the mode
$\mu_1(t)$ at the initial instant ($\mu_1(t_0)\ne 0$), the mode
$\mu_2(t)$ cannot be equal to zero (even if it was zero at
$t=t_0$!) because a {\it nonzero force} $-\frac{\partial
U}{\partial\mu_2}=\gamma\mu_1^2$ appears in the right-hand side of
Eq.(\ref{eq6b}) since $\mu_1(t)\not\equiv 0$. In other words, the
dynamical regime $\mu_1(t)\not\equiv 0, \mu_2(t)\equiv 0$ cannot
exist because of the contradiction with Eq.(\ref{eq6b}). Unlike
this, the dynamical regime $\mu_2(t)\not\equiv 0, \mu_1(t)\equiv
0$ can exist because such a condition does not contradict
equations~(\ref{eq6a}) and~(\ref{eq6b}). We can say that now there
is no force in the right-hand side of Eq.(\ref{eq6b}) because
$\frac{\partial U}{\partial\mu_2}\equiv 0$ as a consequence of the
identity $\mu_1(t)\equiv 0$.

In the last case, the dynamical regime of the  system
(\ref{eq6a},\ref{eq6b}) represents a harmonic oscillation only of
the second variable:
\begin{equation}
\label{eq7}\mu_2(t)=A\cos(\omega_2t+\delta),
\end{equation}
where $A$ and $\delta$ are two arbitrary constants.

Thus, there is {\it force} action from the mode $\mu_1(t)$ on the
mode $\mu_2(t)$, but not vice versa. We proved in~\cite{PhysD}
that such a situation can be realized only in the case where the
symmetry group of the mode $\mu_1(t)$ is less than or equal to
that of the mode $\mu_2(t)$.\footnote{Remember, that speaking
about a mode we must take into account that our time-dependent
coefficient is multiplied by the appropriate basis vector of a
certain  irrep of the group $G$, and namely this vector determines
the symmetry group of the considered mode.}

Nevertheless, the mode $\mu_2(t)$ can excite the mode $\mu_1(t)$
under certain circumstances. Indeed, substituting the
solution~(\ref{eq7}) of Eq.(\ref{eq6b}) into  Eq.(\ref{eq6a}), we
obtain
\begin{equation}
\label{eq8}\ddot\mu_1(t)+\left[\omega_1^2-2A\gamma\cos(\omega_2t+\delta)\right]\mu_1=0.
\end{equation}
By means of simple algebraic transformations this equation can be
converted to the Mathieu equation in its standard form:
\begin{equation}
\label{eq9}z''+\left[a-2q\cos(2\tau)\right]z=0,
\end{equation}
where $z=z(\tau)$. But in the ($a-q$) plane of the Mathieu
equation~(\ref{eq9}), there exist domains of stable and unstable
movement. If pertinent parameters ($\omega_1, \omega_2, \gamma$)
of our dynamical system~(\ref{eq6a},\ref{eq6b}) have values such
that corresponding parameters $a$ and $q$ of Eq.(\ref{eq9}) get
into an unstable domain, then the nonzero function $z(\tau)$ and,
therefore, the mode $\mu_1(t)$ appears. In other words, the
initial dynamical regime  $\mu_1(t)\equiv 0, \mu_2(t)\not\equiv 0$
loses its stability, and a new dynamical regime
$\mu_1(t)\not\equiv 0, \mu_2(t)\not\equiv 0$ arises {\it
spontaneously} for definite values of the parameters of
Eqs.(\ref{eq6a},\ref{eq6b}). Since this phenomenon is similar to
the well-known parametric resonance, we can speak, in such a case,
about {\it parametric action} from the mode $\mu_2(t)$ on the mode
$\mu_1(t)$.

The characteristic property of the parametric interaction is that
the appropriate force ($-\frac{\partial U}{\partial
\mu_1}=2\gamma\mu_1\mu_2$, in our case) vanishes when the mode
($\mu_1$, in our case), on which this force acts, becomes zero.
The following important result was proven in~\cite{PhysD}: the
mode of lower symmetry acts on the mode of higher symmetry by {\it
force} interaction, while the mode of higher symmetry can act on
the mode of lower symmetry only {\it parametrically}.
Consequently, if the parametric excitation of a certain mode does
take place, this phenomenon must by necessity be accompanied by
{\it spontaneous breaking} of symmetry of the mechanical system
vibrational state.

Thus, the initially excited dynamical regime can lose its
stability because of parametric interactions with some zero modes
and, as a result, can transform spontaneously into another
dynamical regime, described by a greater number of dynamical
variables, with appropriate lowering of symmetry. Obviously, we
may treat such phenomenon as a dynamical analog of a phase
transition.

The above mechanism of loss of stability of bushes of vibrational
modes in octahedral structures with Lennard-Jones
interactions will be discussed in Sec.\ref{stability}.

\section{Bushes of vibrational modes for the octahedral structure}
\label{for structure}

In this section, we consider bushes of vibrational modes for the
octahedral structure without a particle in the center of the
octahedron depicted in Fig.~\ref{fig.1}. This mechanical
structure, in its equilibrium state, is described by the point
symmetry group $O_h$, whose brief description and explicit form of
the appropriate irreducible representations can be found in
Appendix 1. The group $O_h$ has ten irreps $\Gamma_j\;
(j=1,\ldots,10)$: four one-dimensional ($\Gamma_1, \Gamma_2,
\Gamma_3, \Gamma_4$), two two-dimensional ($\Gamma_5, \Gamma_6$)
and four three-dimensional. Note that basis
vectors~(\ref{eq20_0}), as well as invariant vectors~(\ref{eq21}),
depend on the explicit form of the matrix irreps $\Gamma_j$ of the
group $G=O_h$,\footnote{Namely because of this reason we give the
explicit form of the irreps of the group $O_h$ in Appendix 1
(remember that a given irrep is determined up to arbitrary unitary
transformation).} but the contributions
$\mbox{\boldmath{$\Delta$}}_j$~(\ref{eq22}) to the vector ${\bf
X}(t)$ from individual $\Gamma_j$ do not depend on the concrete
form of these irreps.

\subsection{Basis vectors of the irreps of the group $O_h$}
\label{Basis vectors of the irreps}

Not all irreps of the group $G$ are contained in the decomposition
of the full mechanical representation $\Gamma_{\rm mech}$ of our system
into its irreducible parts. It can be shown that the irreps
$\Gamma_2, \Gamma_3, \Gamma_4$ and $\Gamma_6$ are not contained in
$\Gamma_{\rm mech}$, while $\Gamma_{10}$ enters into this reducible
representation twice. One of two copies of $\Gamma_{10}$ in the
decomposition of $\Gamma_{\rm mech}$ describes the translational movement of the
mechanical structure as a rigid body. Considering only {\it
vibrational} regimes, we will exclude this copy of $\Gamma_{10}$
from consideration. Because of the same reason we also exclude
the irrep $\Gamma_9$ which describes {\it rotation} of the
considered octahedral structure.\footnote{There exist also
rotational-vibrational bushes for which modes of $\Gamma_9$ turn
to be root modes, but in the present paper, we will study only
pure vibrational bushes.} Thus, there are only $3\times 6-6=12$
pure vibrational degrees of freedom and, as a consequence, we can
restrict ourselves by consideration of the 12-dimensional
vibrational representation $\Gamma_{\rm vibr}$ instead of the
18-dimensional mechanical representation $\Gamma_{\rm mech}$.

Since decomposition of $\Gamma_{\rm vibr}$ contains only one copy
of each irrep $\Gamma_j$ for $j=1,5,7,8,10$, all symmetry adapted
coordinates, generated by the basis vectors of these irreps, turn
out to be the {\it normal} coordinates, and we may not distinguish
these two notations in the further discussion.

We give the basis vectors of the irreps $\Gamma_j$, which form a
basis of the vibrational representation $\Gamma_{\rm vibr}$, in
Table~\ref{1}. These basis vectors were found by the well-known
method of projection operators with the aid of the appropriate
computer program. Each of these 18-dimensional vectors determines
the specific set of displacements $({\bf r}_1|\;{\bf r}_2|\;{\bf
r}_3|\;{\bf r}_4|\;{\bf r}_5|\;{\bf r}_6)$ of the six particles of our
octahedral structure (the numbers of particles are given in
Fig.~\ref{fig.1}). Here ${\bf r}_j=(x_j,y_j,z_j)$ is the
three-dimensional displacement vector of the $j$-th particle from
its equilibrium position determined by its projections along three
Cartesian axes $X, Y, Z$.
\begin{table}[htb]
\caption{Basis vectors of the irreps of the group $O_h$ for the
octahedral structure depicted in Fig.1} \label{1} $$
\begin{array}{|c|c|c|}
\hline
 & \mbox{Basis vectors} &\mbox{Symmetry groups} \\ \hline
\mbox{\boldmath{$\varphi$}}_1^{(1)}&\frac{1}{\sqrt 6}\left(\;
0,0,1\;|\;\mbox-1,0,0\;|\;0,1,0\;|\;1,0,0\;|\;0,\mbox-1,0\;|\;0,0,\mbox-1\;
\right)&O_h\\ \hline
\mbox{\boldmath{$\varphi$}}_1^{(5)}&\frac{1}{\sqrt{12}}\left(
\;0,0,\mbox-2\;|\;\mbox-1,0,0\;|\;0,1,0\;|\;1,0,0\;|\;0,\mbox-1,0\;|\;
0,0,2\;\right)&D_{4h}\\ \hline
\mbox{\boldmath{$\varphi$}}_2^{(5)}&\frac{1}{\sqrt{12}}\left(\;
0,0,0\;|\;\sqrt 3,0,0\;|\;0,\sqrt 3,0\;|\;\mbox-\sqrt
3,0,0\;|\;0,\mbox-\sqrt 3,0\;|\;0,0,0\;\right)&D_{2h}\\ \hline
\mbox{\boldmath{$\varphi$}}_1^{(7)}&\frac{1}{2}\left(\;0,1,0\;|\;0,0,0\;|\;
0,0,1\;|\;0,0,0\;|\; 0,0,\mbox-1\;|\; 0,\mbox-1,0\;
\right)&D'_{2h}\\ \hline
\mbox{\boldmath{$\varphi$}}_2^{(7)}&\frac{1}{2}\left(\;
1,0,0\;|\;0,0,\mbox-1\;|\; 0,0,0\;|\; 0,0,1\;|\;
0,0,0\;|\;\mbox-1,0,0\; \right)&D'_{2h}\\ \hline
\mbox{\boldmath{$\varphi$}}_3^{(7)}&\frac{1}{2}\left(\;
0,0,0\;|\;0,\mbox-1,0\;|\; 1,0,0\;|\; 0,1,0\;|\;\mbox-1,0,0\;|\;
0,0,0\; \right)&D'_{2h}\\ \hline
\mbox{\boldmath{$\varphi$}}_1^{(8)}&\frac{1}{2}\left(\;
1,0,0\;|\;0,0,0\;|\;\mbox-1,0,0\;|\; 0,0,0\;|\;\mbox-1,0,0\;|\;
1,0,0\; \right)&D'_{2d}\\ \hline
\mbox{\boldmath{$\varphi$}}_2^{(8)}&\frac{1}{2}\left(\;
0,\mbox-1,0\;|\;0,1,0\;|\; 0,0,0\;|\; 0,1,0\;|\; 0,0,0\;|\;
0,\mbox-1,0\; \right)&D'_{2d}\\ \hline
\mbox{\boldmath{$\varphi$}}_3^{(8)}&\frac{1}{2}\left(\;
0,0,0\;|\;0,0,\mbox-1\;|\; 0,0,1\;|\; 0,0,\mbox-1\;|\; 0,0,1\;|\;
0,0,0\; \right)&D'_{2d}\\  \hline
\mbox{\boldmath{$\varphi$}}_1^{(10)}&\frac{1}{\sqrt{12}}\left(\;
1,0,0\;|\;\mbox-2,0,0\;|\; 1,0,0\;|\;\mbox-2,0,0\;|\; 1,0,0\;|\;
1,0,0\; \right)&C_{4v}\\ \hline
\mbox{\boldmath{$\varphi$}}_2^{(10)}&\frac{1}{\sqrt{12}}\left(\;
0,1,0\;|\; 0,1,0\;|\; 0,\mbox-2,0\;|\; 0,1,0\;|\; 0,\mbox-2,0\;|\;
0,1,0\; \right)&C_{4v}\\ \hline
\mbox{\boldmath{$\varphi$}}_3^{(10)}&\frac{1}{\sqrt{12}}\left(
0,0,\mbox-2\;|\;0,0,1\;|\; 0,0,1\;|\; 0,0,1\;|\; 0,0,1\;|\;
0,0,\mbox-2\; \right)&C_{4v}\\ \hline
\end{array}
$$
\end{table}
The complete set of basis vectors
$\mbox{\boldmath{$\varphi$}}_i^{(j)}$ from Table~\ref{1} can be
used as a basis for the decomposition of an arbitrary {\it
vibrational} regime ${\bf X}(t)$ (see Eq.(\ref{eq2})).

\subsection{Bushes of vibrational modes}
\label{Bushes of vibrational modes}

As was already discussed in previous sections, a certain symmetry
group $G_D\subseteq G$ corresponds to a given bush. Therefore, to
find all bushes, we must know all {\it subgroups} $G_D^{(k)}$
of the parent group $G=O_h$ of our mechanical system in its
equilibrium state. Moreover, we must take into account not only
different subgroups, but also their different embeddings
(orientations) in the parent group $O_h$. For example, there are
three ways of embedding of the group $G_D=C_{2v}$ (we denote them
by primes on the appropriate Schoenflies symbols:
$C'_{2v}, C''_{2v}$) -- they differ from each other by their
generators. We give the generators of all subgroups of the group
$O_h$, corresponding to different nonlinear vibrational regimes,
in the second column of Tables~\ref{2a},\ref{2b}. For the above
case we see the following generators for three variants of the
subgroup $C_{2v}$:
\begin{equation}
\label{eq30_1} C_{2v}: h_4, h_{26};\quad
C'_{2v}:h_{13},h_{28};\quad C''_{2v}:h_4,h_{37}.
\end{equation}
Using the list of symmetry elements from~\cite{Kovalev}, one can
find that in $C_{2v}$ the two-fold axis $h_4$ is the coordinate
axis $C_2^z$ and the mirror plane $h_{26}$ is the coordinate plane
$\sigma_x$; in $C'_{2v}$, $h_{13}$ is the diagonal two-fold axis
$C_2^{\bar xy}$, while the mirror plane $h_{28}$ is the coordinate
plane $\sigma_z$; in $C''_{2v}$, $h_4$ is the coordinate axis
$C_2^z$ and the mirror plane $h_{37}$ coincides with the diagonal
plane $\sigma_{\bar xy}$.

All subgroups $G_D^{(k)}$ of the group $G=O_h$ with their
different embeddings and the sets of appropriate invariant vectors
for each irrep of this group can be found
in~\cite{Koptsik}.\footnote{The invariant vectors of irreps of
group $O_h$ were obtained in the paper~\cite{Koptsik} in
connection with discussion of some problems of the theory of color
symmetry.}

Let us remember, that for obtaining a given bush of vibrational
modes with symmetry group $G_D\subseteq G$, we must solve the
linear algebraic equations~(\ref{eq5}) for each of five irreps
$\Gamma_j\;(j=1,5,7,8,10)$ of the group $G=O_h$. In such a  way,
we obtain the invariant vector $\mbox{\boldmath{$\mu$}}_j$
individually for every irrep $\Gamma_j$. It is very essential that
for many irreps these invariant vectors turn out to be {\it zero
vectors}. For example, for $G_D=D_{4h}$ there are only two irreps
to which nonzero invariant vectors correspond: the one-dimensional
irrep $\Gamma_1$ and the two-dimensional irrep $\Gamma_5$. Moreover,
the invariant vector for $\Gamma_5$ has a special form:
$\mbox{\boldmath{$\mu$}}_5=(a,0)$, where $a$ is an arbitrary
constant appearing in the solution of Eq.(\ref{eq5}). Therefore, we obtain
the following expression\footnote{A reader can see this result in
Table~\ref{2a}.} for the bush $B2[D_{4h}]$, if replacing the
arbitrary constants by time-dependent functions $\mu_i^{(j)}(t)$:
\begin{equation}
\label{eq31_1} {\bf
X}(t)=\mu_1^{(1)}(t)\mbox{\boldmath{$\varphi$}}_1^{(1)}+
\mu_1^{(5)}(t)\mbox{\boldmath{$\varphi$}}_1^{(5)}.
\end{equation}
(Note that the  coefficient before the second basis vector
$\mbox{\boldmath{$\varphi$}}_2^{(5)}$ of the irrep $\Gamma_5$ is
equal to zero!). Analogously, for bush $B3[D'_{2d}]$, we find that
only three irreps $\Gamma_1,\Gamma_5$ and $\Gamma_8$ contribute to
the corresponding dynamical regime:
\begin{equation}
\label{eq32_1} {\bf
X}(t)=\mu_1^{(1)}(t)\mbox{\boldmath{$\varphi$}}_1^{(1)}+\mu_1^{(5)}(t)\mbox{\boldmath{$\varphi$}}_1^{(5)}+\mu_3^{(8)}(t)\mbox{\boldmath{$\varphi$}}_3^{(8)}.
\end{equation}
Again, we see that a very specific invariant vector
$\mbox{\boldmath{$\mu$}}_8=(0,0,a)$ corresponds to the
three-dimensional irrep $\Gamma_8$.

Comparing Eqs.(\ref{eq31_1}) and (\ref{eq32_1}), one finds the
identical symbols $\mu_1^{(1)}(t)$ and $\mu_1^{(5)}(t)$ in two
different bushes $B2[D_{4h}]$ and $B3[D_{2d}]$. There is no
connection between  these functions, because we exploit here, as
well as in Tables~\ref{2a},\ref{2b}, the following convention: one
and the same notation $\mu_i^{(j)}(t)$ corresponds to {\it
different} functions when it is used in the description of
different bushes.

Let us consider Tables~\ref{2a} and \ref{2b} where all bushes of
vibrational modes for the octahedral mechanical structure are
listed. The root modes of bushes are {\it underlined} (remember,
that root mode possesses the minimal symmetry among all other
modes contained in the given bush)\footnote{A more exact
discussion of this topic see below.}.

Similar to phase transition theory, we can speak about dynamical
domains for a given bush. Indeed, as it can be seen from
Tables~\ref{2a},\ref{2b}, there is only one bush with the symmetry
group $G_D=D_{4h}$, namely, the bush $B2[D_{4h}]$. The 4-fold axis
of this group, associated with the generator $h_{14}$, is directed
along the $Z$ coordinate axis, i.e., $G_D=D^z_{4h}$ (see
geometrical sense of the generators from the third column of
Table~\ref{2a} in the handbook~\cite{Kovalev}). Obviously, there
must exist also bushes with the same symmetry group $D_{4h}$, but
with 4-fold axis oriented along $X$ and $Y$ coordinate axis, i.e.
$B[D^x_{4h}]$ and $B[D^y_{4h}]$, because of the equal status of
the $X,Y,Z$ axes in the parent group $G=O_h$. The collection of
these three {\it equivalent}\footnote{Two bushes are equivalent to
each other, if there exist a symmetry element $g$ of the parent
group $G$ which transforms one bush to another.} bushes is similar
to the set of different domains in the phase transition
theory\footnote{The total number of such domains is equal to the
index of the subgroup $G_D$ in the group $G$, i.e., to the ratio
of the orders of these two groups: $\|G\|/\|G_D\|$.}. In
Tables~\ref{2a},\ref{2b} we include only one bush from the
complete collection of the equivalent bushes.

As it can be seen from Tables~\ref{2a},\ref{2b}, the root modes
corresponding to some bushes represent superpositions of several
basis vectors of a given multidimensional irrep. For example, we
find in the fifth column of this table that the root modes for
bushes $B7[D_3], B8[C_{3v}]$ and $B9[D_{3d}]$ are simply sums of
all three basis vectors of the irreps $\Gamma_8,\Gamma_{10}$ and
$\Gamma_7$, respectively. In terms of invariant vectors, this
means that in solving Eq.(\ref{eq5}) we obtain for the above
irreps the solutions of the same form:
$\mbox{\boldmath{$\mu$}}_8=\mbox{\boldmath{$\mu$}}_{10}=\mbox{\boldmath{$\mu$}}_7=(a,a,a)$,
where $a$ is an arbitrary constant (this constant is different for
each of the irreps $\Gamma_8,\Gamma_{10}$ and $\Gamma_7$!). In writing
the bushes  $B7[D_3], B8[C_{3v}]$ and $B9[D_{3d}]$ we must replace
these constants by the appropriate functions of time
$\mu_1^{(8)}(t), \mu_1^{(10)}(t)$ and $\mu_1^{(7)}(t)$, respectively.

Let us now consider the more complicated cases where the choice of
root mode for a given bush is {\it not uniquely defined}. For
example, there are two different variants for choosing the root
mode for bush  $B6[C'_{2v}]$: either the sum
$(\mbox{\boldmath{$\varphi$}}_1^{(8)}+\mbox{\boldmath{$\varphi$}}_2^{(8)})$
of two basis vectors of three-dimensional irrep $\Gamma_8$, or
the difference
$(\mbox{\boldmath{$\varphi$}}_1^{(10)}-\mbox{\boldmath{$\varphi$}}_2^{(10)})$
of two basis vectors of another three-dimensional irrep
$\Gamma_{10}$ can represent the root mode.\footnote{The invariant
vectors $(a,a,0)$ and $(a,\mbox -a,0)$ correspond to these two
cases.} Indeed, the excitation of each of these modes {\it
separately} or {\it simultaneously} leads to the excitation of the
same bush $B6[C'_{2v}]$.

Such a possibility of different choices of the root mode is brought
about by the fact that both combinations of the basis vectors
$(\mbox{\boldmath{$\varphi$}}_1^{(8)}+\mbox{\boldmath{$\varphi$}}_2^{(8)})$
and
$(\mbox{\boldmath{$\varphi$}}_1^{(10)}-\mbox{\boldmath{$\varphi$}}_2^{(10)})$
possess the same symmetry group  $G_D=C'_{2v}$, while the basis
vectors of other irreps contributing to bush  $B6[C'_{2v}]$ have
{\it higher} symmetry, namely,
$\mbox{\boldmath{$\varphi$}}_1^{(1)}\mbox{ -- } O_h$,
$\mbox{\boldmath{$\varphi$}}_1^{(5)}\mbox{ -- } D_{4h}$,
$\mbox{\boldmath{$\varphi$}}_3^{(7)}\mbox{ -- } D'_{2h}$ (these
three point groups are supergroups with respect to the symmetry
group $C'_{2v}$ of the root modes). From this example we see that
all modes of a given bush do have symmetry higher than or equal to
that of the root mode (see Sec.~\ref{Geometrical aspects}).

Above, we consider the cases where bushes may be excited by the
initial excitation of a {\it single} mode (even if there are
different variants of the choice of this single root mode). But
there exists bushes of another type, namely, the bushes whose
excitation is possible only if {\it two different modes}, i.e.
modes belonging to the different irreps, are excited {\it
simultaneously}. For the considered mechanical system there are
four bushes of such a type in Table~\ref{2b}:
\begin{equation}
\label{39} B11 [D'_2] \mbox{ -- } \Gamma_7, \Gamma_8;\; B12
[C_{2v}] \mbox{ -- } \Gamma_8, \Gamma_{10};\; B13 [C''_{2v}]
\mbox{ -- } \Gamma_7, \Gamma_{10};\; B17 [C_3] \mbox{ -- }
\Gamma_8, \Gamma_{10}.
\end{equation}
Here we also point out the pairs of the irreps whose modes must be
excited for the excitation of the corresponding bushes.

Let us consider the bush \(B13 [C''_{2v}]\) in more detail. We can
write down the symmetry groups of all four modes contributing to
it (these groups can be found in Table~\ref{1}):
\begin{equation}
\label{40} \mbox{\boldmath{$\varphi$}}_1^{(1)}\mbox{ -- } O_h,\;
\mbox{\boldmath{$\varphi$}}_1^{(5)}\mbox{ -- } D_{4h},\;
\mbox{\boldmath{$\varphi$}}_3^{(7)}\mbox{ -- } D'_{2h},\;
\mbox{\boldmath{$\varphi$}}_3^{(10)}\mbox{ -- } C_{4v}.
\end{equation}
Note that the symmetry group $C''_{2v}$ of the whole
bush \(B13 [C''_{2v}]\) is not found among the symmetry groups (\ref{40}) of
its individual modes. This fact can be understood as follows.
Since our bush is the superposition of four modes
$\mbox{\boldmath{$\varphi$}}_1^{(1)}$,$\mbox{\boldmath{$\varphi$}}_1^{(5)}$,
$\mbox{\boldmath{$\varphi$}}_3^{(7)}$ and
$\mbox{\boldmath{$\varphi$}}_3^{(10)}$, it is clear that its
symmetry group must be the {\it intersection} of the four symmetry
groups corresponding to these modes:
\begin{equation}
\label{41} O_h(h_2,h_3,h_5,h_{13},h_{25}),
D_{4h}(h_2,h_{14},h_{25}), D'_{2h}(h_4,h_{13},h_{25}),
C_{4v}(h_{14},h_{37}).
\end{equation}
Here we write the generators of each group in the parentheses next to
its Schoenflies symbol (these generators can be found in the third
column of Table~\ref{2a} near the appropriate symbol of the symmetry
group)\footnote{Only for group $O_h$, we use the extended set of
generators.}.

The two first groups from (\ref{41}) --- $O_h$ and $D_{4h}$ ---
are subgroups with respect to of the both last groups, $D'_{2h}$
and $C_{4v}$: all elements of $D'_{2h}$ and $C_{4v}$ are contained
in $D_{4h}$ and therefore also in $O_h$. Indeed, the group
$D_{4h}(h_2,h_{14},h_{25})$ contains four two-fold horizontal axes
(rotations $h_2,h_3,h_{13},h_{17}$) and four vertical mirror
planes (reflections $h_{26},h_{27},h_{37},h_{41}$) intersecting
along vertical four-fold axis (rotations $h_4,h_{14},h_{15}$)
coinciding with the $Z$ coordinate axis, and it is obvious that
the generators of both groups $D'_{2h}$ and $C_{4v}$ are identical
with some elements of the group
$D_{4h}$\footnote{\label{ftnt13}Using the multiplication table
from \cite{Kovalev} for the element of the group $O_h$, we can
obtain the generators of $D'_{2h}$ and $C_{4v}$ from the
generators of $D_{4h}$ as follows: \(h_4=h_{14}^2,h_{13}=h_2
h_{14}, h_{37}=h_{13} h_{25}\).}.

Let us now compare the elements of the groups
\(D'_{2h}(h_4,h_{13},h_{25})\) and  \(C_{4v}(h_{14},h_{37})\). The
second generator of $C_{4v}$ is contained in $D'_{2h}$ since
$h_{37}=h_{13}\cdot h_{25}$, and the first generator of $D'_{2h}$
is contained in $C_{4v}$ since $h_{14}^2=h_4$ (see footnote
\ref{ftnt13}). Then we can write the {\it extended} sets of the
generators of the groups $C_{4v}$ and $D'_{2d}$ as follows:
\(C_{4v}(h_{14},\underline{h_4,h_{37}})\),
\(D'_{2d}(h_{13},h_{25},\underline{h_4,h_{37}})\). The
intersection of these two groups is a group with their common
elements, i.e., the group with the generators $h_4$ and $h_{37}$.
In our notation, this is the group \(C''_{2v}(h_4,h_{37})\)
associated with the whole bush \(B13 [C''_{2v}]\) in
Table~\ref{2b}.

Other three bushes from the list (\ref{39}) can be considered
similarly to the above discussed bush  \(B13 [C''_{2v}]\).

Unlike the above cases, the modes of only one irrep are root modes
for bushes
\begin{equation}
\label{42} B10 [D_{2h}], B14 [C'_2], B16 [C'_s], B18 [C_i],
\end{equation}
but these modes are determined by {\it several} independent
coefficients $\mu_i^{(j)}(t)$. For example, the root mode
\(\mu_1^{(5)}(t)\mbox{\boldmath{$\varphi$}}_1^{(5)}+\mu_2^{(5)}(t)\mbox{\boldmath{$\varphi$}}_2^{(5)}\)
corresponds to the bush $B10 [D_{2h}]$ (This mode is generated by the
invariant vector of the form $(a,b)$, associated with the
two-dimensional irrep $\Gamma_5$). Analogously, the root mode
\(\mu_1^{(10)}(t)
(\mbox{\boldmath{$\varphi$}}_1^{(10)}+\mbox{\boldmath{$\varphi$}}_2^{(10)})+
\mu_3^{(10)}(t)\mbox{\boldmath{$\varphi$}}_3^{(10)}\) corresponds
to the bush $B16 [C'_s]$ (this mode is induced by the invariant
vector of the form $(a,a,b)$ of the three-dimensional irrep
$\Gamma_{10}$).

\section{Exact dynamical equations for bushes of vibrational modes}
\label {exact}

We consider a mechanical system  of six mass points (particles)
whose interactions are described by a pair isotropic potential
$u(r)$ where $r$ is the distance between two particles. We suppose
that in the {\it equilibrium state} these particles form a regular
octahedron with edge $a_0$ which is depicted in Fig.~\ref{fig.1}.
Let us introduce a Cartesian coordinate system. Four particles of
the above octahedron lie in the $XY$ plane and form a square with
edge $a_0$. Two other particles lie on the $Z$ axis and we will speak
about the ``top particle" and the ``bottom particle" with respect
to the direction of the $Z$ axis. Obviously, the distance between
each of these two particles and any of the four particles in the
$XY$ plane is equal to $a_0$.

In the equilibrium state, our mechanical system possesses the
point symmetry group $O_h$. All possible bushes of vibrational
modes were considered for this system in the previous section, and
we remember that a certain subgroup $G_D$ of the group $G=O_h$
corresponds to each of these bushes.

In this section, we consider $u(r)$ as an {\it arbitrary} pair
isotropic potential, but for studying the bush stability in
Sec.\ref{stability} we suppose that $u(r)$ is the well-known
Lennard-Jones potential:
\begin{equation}
\label{eq10} u(r)=\frac{A}{r^{12}}-\frac{B}{r^6}.
\end{equation}
Here $A$ and $B$ are certain constants characterizing the value of the
repulsive and attractive potential parts, respectively.

The potential energy of our system in its vibrational state can be
written in the form
\begin{equation}
\label{eq11} V({\bf X})=\sum_{{i,j}\atop {\left( i<j \right) }}
u(r_{ij}),
\end{equation}
where $r_{ij}$ is a distance between the $i$-th and $j$-th particles. The
$N$-dimensional vector ${\bf X}=\left( x_1(t),
x_2(t),\ldots,x_N(t)\right)$ in Eq.~(\ref{eq11}),
already introduced in Sec.\ref{bushes}, determines the displacements
of all particles at an arbitrary chosen instant $t$. Note that in
our present case $N=18$.\footnote{In this section, we denote by
$N$ the total number of degrees of freedom of the considered
mechanical system (in other sections, $N$ is the number of
particles).}

Since all particles possess identical masses, which we suppose to
be equal to unity, the dynamical equations of Newton type can be
written as follows:
\begin{equation}
\label{eq12} \ddot x_i=-\frac {\partial V}{\partial x_i},\quad
i=1,\ldots,N \qquad (N=18).
\end{equation}
It is easy to find such a scaling transformation that both
constants $A$ and $B$ in the potential energy $V(\bf X)$ (see
Eqs.(\ref{eq10}, \ref{eq11}) will be equal to unity $(A=1, B=1)$:
\begin{equation}
\label{eq13}
\begin{array}{l}
 x_i=\left( \frac{A}{B}\right)^{\frac{1}{6}}\tilde x_i,
\\ t=\frac{m^\frac{1}{2}A^\frac{2}{3}}{B^\frac{7}{6}}\tilde t.
\end{array}
\end{equation}
Because of this transformation, there are no arbitrary constants
in Eq.(\ref{eq12}) and, therefore, the results obtained in the
present paper are universal with respect to all Lennard-Jones
potentials. Supposing that the scaling transformation (\ref{eq13}) is
already done, we will not change the notation of the time variable
$t$ and space variables $x_i$ in the further discussion.

\subsection{Lagrange equations for bush modes}
\label{Lagrange equations}

For obtaining dynamical equations of bushes of modes, we must
transfer from the original (old) variables $x_i(t)\;
(i=1,2,\ldots,N)$ to new variables $\mu_j(t)\;(j=1,2,\ldots,N)$,
which describe the amplitudes of modes in the decomposition
(\ref{eq2})
\begin{equation}
\label{eq20} {\bf X}(t)=\sum_{j=1}^N \mu_j(t)
{\mbox{\boldmath$\varphi$}}_j.
\end{equation}
In contrast to Eq.(\ref{eq2}) we use here a single index $j$ for numbering the
basis vectors of all irreps of the group $G$.

When a certain bush with a symmetry group  $G_D \subset G$ is
considered, some amplitudes $\mu_j(t)$ turn out to be equal to
zero ($\mu_j(t)\equiv 0$) and we can suppose without loss of
generality that index $j$ from Eq.(\ref{eq20}) runs over only the
first $n$ values ($j=1,2,\ldots,n$). Here $n<N$ for all bushes of
modes except for the bush with trivial symmetry group $G_D=1$.
Thus, a given bush represents a dynamical system whose dimension
is {\it less} than that of the original mechanical system ($n<N$)
and its time evolution can be described with the aid of $n$
generalized variables $\mu_j(t)\;(j=1,\ldots,n)$. The Lagrange method
seems to be the most natural and convenient for obtaining the
dynamical equations for bushes of modes. Let us consider this idea
in more detail.

For a given $n$-dimensional bush we can rewrite Eq.(\ref{eq20}) in
the form
\begin{equation}
\label{eq30} {\bf X}(t)=\sum_{j=1}^n \mu_j(t)
{\mbox{\boldmath$\varphi$}}_j.
\end{equation}
Since all basis vectors ${\mbox{\boldmath$\varphi$}}_j$ are known,
Eq.(\ref{eq30}) provides explicit expression of each of the $N$ old
variables $x_i(t)\; (i=1,\ldots,N)$ as a function (linear
superposition!) of $n$ new variables $\mu_j(t)\; (j=1,\ldots,n)$.
Therefore, the potential energy $\widetilde V(\mu_1,\ldots,\mu_n)$
in the new variables can be obtained from the potential energy
$V(x_1,\ldots,x_N)$ in the old variables by substitution $x_i(t)\;
(i=1,\ldots,N)$ from Eq.(\ref{eq30}) to the function
$V(x_1,\ldots,x_N)$. Similarly, the kinetic energy $T(\dot
x_1,\ldots,\dot x_N)=\frac{1}{2}\sum_{i=1}^N \dot x_i^2$ can be
transformed with the aid of Eq.(\ref{eq30}) to the following
quadratic form in terms of velocities $\dot \mu_j(t)$ of the new
variables:
\begin{equation}
\label{eq31}\widetilde T(\dot \mu_1,\ldots,\dot
\mu_n)=\frac{1}{2}\sum_{j=1}^n\sum_{k=1}^n \dot \mu_j\dot \mu_k
\langle
{\mbox{\boldmath$\varphi$}}_j|{\mbox{\boldmath$\varphi$}}_k\rangle.
\end{equation}
Here $\langle
{\mbox{\boldmath$\varphi$}}_j|{\mbox{\boldmath$\varphi$}}_k\rangle$
is the scalar product of two {\it real} $N$-dimensional basis
vectors ${\mbox{\boldmath$\varphi$}}_j$ and
${\mbox{\boldmath$\varphi$}}_k$. We can suppose that the basis
vectors ${\mbox{\boldmath$\varphi$}}_j \; (j=1,\ldots,N)$ are
orthonormal:
\begin{equation}
\label{eq32} \langle
{\mbox{\boldmath$\varphi$}}_j|{\mbox{\boldmath$\varphi$}}_k\rangle=\delta_{jk}.
\end{equation}
(Note that basis vectors from Table~\ref{1} do satisfy this
condition).

Taking into account Eq.(\ref{eq32}), we can rewrite
Eq.(\ref{eq31}) as follows:
\begin{equation}
\label{eq33}\widetilde T(\dot \mu_1,\ldots,\dot
\mu_n)=\frac{1}{2}\sum_{j=1}^n \dot \mu_j^2.
\end{equation}
Then with the aid of the Lagrange function $\widetilde
L=\widetilde T - \widetilde V$, the exact dynamical equations for
a given $n$-dimensional bush with the symmetry group $G_D \subset
G$ can be obtained:
\begin{equation}
\label{eq34}\frac{d}{dt}\left( \frac{\partial \widetilde L
}{\partial \dot \mu_j}\right)-\frac{\partial \widetilde
L}{\partial \mu_j}=0 \qquad (j=1,\ldots,n).
\end{equation}

Let us emphasize that these dynamical equations of the bush of
modes are {\it exact}. Actually, this fact is brought about by the
symmetry-related restriction
\begin{equation}
\label{eq35} \widehat G_D {\bf X}(t)={\bf X}(t)
\end{equation}
which reduces the number of degrees of freedom from $N$ to $n$.
Indeed, namely the condition (\ref{eq35}) leads to the vanishing
of some terms in  Eq.(\ref{eq20}) and reduces it to
Eq.(\ref{eq30}).

\subsection{Examples of dynamical equations for bushes of vibrational modes}
\label{Examples of dynamical equations}

The exact equations (\ref{eq34}) for bushes of vibrational modes
for the octahedral structure with the Lennard-Jones potential are
rather complicated. Because of this reason, we wrote a special
MAPLE-program for obtaining these equations and for studying the
stability of bushes of modes in modal space (see
Sec.~\ref{mu-space}). Let us now write down the very clear and
compact forms of the exact dynamical equations for some bushes in
terms of natural {\it geometrical variables}.

We consider the following three bushes from
Table~\ref{2a}: $B1[O_h], B2[D_{4h}]$ and $B4[C_{4v}]$, whose
dimension is equal to 1, 2 and 3, respectively. Note that symmetry
groups of the above bushes are connected to each other by the
following group-subgroup relation: \( C_{4v}\subset D_{4h}\subset
O_h. \)

The geometrical forms of our mechanical system in the {\it
vibrational state}, corresponding to the these bushes, can be
revealed from the superpositions of modes (given in the fifth
column of Table~\ref{2a}). But it is easier to understand the
geometrical sense of the above bushes immediately from the
symmetry groups corresponding to them.

Indeed, the one-dimensional bush $B1[O_h]$ consists of only one
(``breathing") mode: the appropriate nonlinear dynamical regime
${\bf X}(t)= \mu_1^{(1)}(t)\mbox{\boldmath{$\varphi$}}_1^{(1)}$
describes evolution of a regular octahedron whose edge $a=a(t)$
periodically changes in time.

The two-dimensional bush $B2[D_{4h}]$ describes a dynamical regime
with two degrees of freedom: ${\bf X}(t)=
\mu_1^{(1)}(t)\mbox{\boldmath{$\varphi$}}_1^{(1)}+
\mu_1^{(5)}(t)\mbox{\boldmath{$\varphi$}}_1^{(5)}$. The symmetry
group $G_D=D_{4h}$ of this bush contains the 4-fold axis
coinciding with the $Z$ coordinate axis and the mirror plane
coinciding with the $XY$ plane. This symmetry group restricts
essentially the form of the polyhedron describing our mechanical
system in the vibrational state. Indeed, the presence of the
4-fold axis demands that the quadrangle in the $XY$ plane be a square.
Because of the same reason, the four edges connecting the particles in
the $XY$ plane (vertices of the above square) with the top
particle lying on the $Z$ axis must be of the same length which we
denote by $b(t)$.

Similarly, let the length of the edges connecting the bottom
particle on the $Z$ axis with any of the 4 particles in the $XY$ plane be
denoted by $c(t)$. In our present case of the bush $B2[D_{4h}]$,
$b(t)=c(t)$ for any time $t$ because of the presence of the
horizontal mirror plane in the group $G_D=D_{4h}$. But for the
three-dimensional bush $B4[C_{4v}]$, described by ${\bf X}(t)=
\mu_1^{(1)}(t){\mbox{\boldmath$\varphi$}}_1^{(1)}
+\mu_1^{(5)}(t){\mbox{\boldmath$\varphi$}}_1^{(5)}
+\mu_3^{(10)}(t){\mbox{\boldmath$\varphi$}}_3^{(10)}$, this mirror
plane is absent and, therefore, $b(t)\ne c(t)$.

Let us also introduce two heights, $h_1(t)$ and $h_2(t)$,
corresponding to the perpendiculars dropped, respectively, from
the top and bottom vertices of our polyhedron onto the $XY$ plane.
Now we can write the dynamical equations of the above bushes in
terms of pure geometrical variables $a(t), b(t), c(t), h_1(t)$ and
$h_2(t)$.

We choose $a(t)$ and $h(t)\equiv h_1(t)\equiv h_2(t)$ as dynamical
variables for describing the two-dimensional bush $B2[D_{4h}]$ and
$a(t), h_1(t)$ and $h_2(t)$ as dynamical variables for describing
the three-dimensional bush $B4[C_{4v}]$. Using these variables we
can write down the potential energy for our bushes of vibrational
modes as follows: $$
\begin{array}{lcl}
B1[O_h]&:& V_{B1}(a)=12u(a)+3u(\sqrt{2}\,a),\\
B2[D_{4h}]&:&V_{B2}(a,h)=4u(a)+2u(\sqrt{2}\,a)+8u\left(\sqrt{h^2+\frac{a^2}{2}}\,\right)
+u(2h),\\
B4[C_{4v}]&:&V_{B4}(a,h_1,h_2)=4u(a)+2u(\sqrt{2}\,a)+4u(b)+4u(c)+u(h_1+h_2),
\end{array}
$$

where
$b=\sqrt{\frac{a^2}{2}+\left(\frac{5}{4}h_1-\frac{1}{4}h_2\right)^2}$,
$c=\sqrt{\frac{a^2}{2}+\left(\frac{5}{4}h_2-\frac{1}{4}h_1\right)^2}$.

Then with the aid of the Lagrange method, we can obtain the
following dynamical equations for the above bushes of vibrational
modes:
\begin{equation} \label{ur}
\begin{array}{ll}
B1[O_h]:&\\ \qquad \ddot a= & -4u'(a)-\sqrt{2}u'(\sqrt{2}a); \\
B2[D_{4h}]:& \\ \qquad \ddot a= &
-2u'(a)-\sqrt{2}u'(\sqrt{2}a)-2u'(b)\frac{a}{b}, \\ \qquad \ddot
h= & -4u'(b)\frac{h}{b}-u'(2h); \\ B4[C_{4v}]:& \\ \qquad \ddot a=
& -2u'(a)-\sqrt{2}u'(\sqrt{2}a)-u'(b)\frac{a}{b}-u'(c)\frac{a}{c},
\\ \qquad \ddot h_1= & -u'(b)\frac{5h_1-h_2}{b}-u'(h_1+h_2), \\
 \qquad \ddot h_2= & -u'(c)\frac{5h_2-h_1}{c}-u'(h_1+h_2).
\end{array}
\end{equation}

Thus, we obtain the dynamical equations of our bushes of
vibrational modes in terms of variables with explicit geometrical
sense. Each bush describes a certain nonlinear dynamical regime
corresponding to such a vibrational state of the considered
mechanical system, that at any fixed time the configuration of
this system is represented by a definite polyhedron with symmetry
group $G_D$ of the given bush.

We can write dynamical equations for the above bushes in terms of
vibrational modes as well. In spite of the more complicated form,
these equations turn out to be more useful for the bush theory,
since they allow us to decompose the appropriate nonlinear
dynamical  regimes into modes of different importance for the case
of small oscillations -- root modes and secondary modes of
different orders~\cite{PhysD}. As an example, we write below the
dynamical equations for the bush $B4[C_{4v}]$ in terms of its
three modes, $\mu_3^{(10)}(t)\equiv \gamma(t)$ (root mode),
$\mu_1^{(1)}(t)\equiv\mu(t)$, $\mu_1^{(5)}(t)\equiv\nu(t)$
(secondary modes):
\begin{eqnarray}
\ddot{\mu}&=&-\frac{1}{6}(4\sqrt{2}u'(a)+4u'(\sqrt{2}a)+2u'(h_1+h_2)+
\frac{u'(b)}{b}(2\sqrt{2}a+5h_1-h_2)+ \nonumber \\ & &
\frac{u'(c)}{c}(2\sqrt{2}a+5h_2-h_1)), \label{200a}\\
\ddot{\nu}&=&-\frac{1}{6}(2\sqrt{2}u'(a)+2u'(\sqrt{2}a)-2u'(h_1+h_2)+
\frac{u'(b)}{b}(\sqrt{2}a-5h_1+h_2)+ \nonumber \\ & &
\frac{u'(c)}{c}(\sqrt{2}a-5h_2+h_1)), \label{200b}\\
\ddot{\gamma}&=&\frac{1}{4}\left(u'(b)\frac{5h_1-h_2}{b}-u'(c)\frac{5h_2-h_1}{c}\right).
\label{200c}
\end{eqnarray}
Here
\begin{equation}\label{201}
\begin{array}{c}
a=\sqrt{2}(r_0+\mu+\nu);
b=\sqrt{(r_0+\mu+\nu)^2+(r_0+\mu-2\nu-3\gamma)^2};\\
c=\sqrt{(r_0+\mu+\nu)^2+(r_0+\mu-2\nu+3\gamma)^2};\\
h_1=r_0+\mu-2\nu-2\gamma; h_2=r_0+\mu-2\nu+2\gamma;
r_0=\frac{a_0}{\sqrt{2}}.
\end{array}
\end{equation}

Let us stress that all above equations are valid not only for the
Lennard-Jones potential (\ref{eq10}), but for any pair and
isotropic potential $u(r)$, as well.

Now, we want to demonstrate that
Eqs.(\ref{200a},\ref{200b},\ref{200c}) possess the specific bush
structure which is discussed in Sec.\ref{bushes}. Note that there
are the definite relations between the symmetry groups of the
modes $\mu(t), \nu(t)$ and $\gamma(t)$:
\begin{equation}
\label{202} G_D[\mu]=O_h \supset G_D[\nu]=D_{4h} \supset
G_D[\gamma]=C_{4v}
\end{equation}
(see Tables~\ref{1} and~\ref{2a}). Therefore, it follows from the
general bush theory that the mode $\gamma(t)$, being of minimal
symmetry, must provide some forces in the right-hand side (rhs) of
the Eqs.(\ref{200a}) and (\ref{200b}) {\it even if} $\mu(t)\equiv
0, \nu(t)\equiv 0$ and, analogously, the mode $\nu(t)$ must
provide a certain force only in the rhs of Eq.(\ref{200a}) when
$\mu(t)\equiv 0$ and $\gamma(t)\equiv 0$. These properties can be
easily seen from Eqs.(\ref{200a}--\ref{201}). One can also reveal
that the mode $\mu(t)$, whose symmetry is higher than those of the
modes $\nu(t)$ and $\gamma(t)$, does not produce any forces in the
rhs of Eqs.(\ref{200b}) and (\ref{200c}), if $\nu(t)\equiv 0$ and
$\gamma(t)\equiv 0$. Similarly, one can verify that the modes
$\nu(t)$ and $\mu(t)$ do not produce any forces in the rhs of
Eq.(\ref{200c}) when $\mu(t)\not\equiv 0, \nu(t)\not\equiv 0$, but
$\gamma(t)\equiv 0$.

Thus, we confirmed the typical bush structure
of the dynamical equations (\ref{200a}--\ref{200c}) in terms of
the individual modes.

\section{Stability of bushes of vibrational modes}
\label{stability}

All possible bushes of vibrational modes for the octahedral
structure depicted in Fig.~\ref{fig.1} were found by means of
group-theoretical methods in the previous section, and these results
are independent of the type of interactions between individual
particles of our mechanical system. On the other hand, studying
the stability of bushes of modes depends essentially on the
concrete type of interactions in the considered system, and we
suppose that they can be described by the Lennard-Jones potential
(\ref{eq10}). Then the bush stability can be analyzed with the aid
of numerical methods.

\subsection{Setting up the problem}
\label{set_prob}

First of all, we must define what we mean speaking about stability
of bushes of vibrational modes, since the term ``stability" is used
in very different senses. Let us excite a given bush $B[G_D]$
with the aid of the appropriate {\it initial conditions} (see
below). We can speak about the energy $E_0$ of the mechanical
system in this initial state. Remember that the given bush is a
closed dynamical object, i.e. energy $E_0$ can spread only among
its own modes. Let us decompose the $3\times N$ -dimensional
vector ${\bf X}(t)$, describing our bush $B[G_D]$ at a certain
instant $t$ into the complete set of orthonormal basis vectors from
Table~\ref{1}. As a result, we obtain a definite collection of
nonzero coefficients $\mu_i^{(j)}(t)$ from Eq.(\ref{eq2}).
We can repeat this procedure at other times during the evolution of
the system, thereby obtaining the time evolution of $\mu_i^{(j)}(t)$.
During this time evolution, $\mu_i^{(j)}(t)$ may pass through
zero but will never be {\it identically} equal to zero.  This set of modes
forms the excited bush $B[G_D]$ of vibrational modes.

Increasing gradually the energy $E_0$ of the initial excitation,
we can detect the loss of stability of the bush $B[G_D]$. Indeed,
beyond a certain value of $E_0$, after some transient time
interval, the complete set of modes with nonzero amplitudes
$\mu_i^{(j)}(t)$ becomes {\it larger} than that of the bush
$B[G_D]$. Obviously, this phenomenon leads to the appearance of a
certain {\it new bush} $\tilde{B}[\widetilde{G_D}]$ which includes
the old bush $B[G_D]$ and whose symmetry $\widetilde{G_D}$ is {\it
lower} than the symmetry group $G_D$ of the bush $B[G_D]$
$(\widetilde{G_D} \subset G_D)$, because all nonzero vibrational
modes with symmetry higher than or equal to $G_D$ are already
contained in $B[G_D]$.

Thus, the loss of stability of a given bush is accompanied by
the spontaneous breaking of symmetry of the initial excited dynamical
regime, described by this bush. We already discussed this
phenomenon  in Sec.\ref{bushes} and concluded that its cause is
analogous to that of parametric resonance.

We can also say this in other words. A given bush $B[G_D]$ represents a
certain dynamical regime in the considered mechanical system. Its
modes interact with other modes which do not belong to $B[G_D]$,
but these interactions must be of parametric (not force!) type
only (see Sec.\ref{Dynamical aspects}). For the appropriate
initial conditions we can get into a region of unstable movement.
As a result, some new modes are excited which were forbidden by the
principle of determinism of classical mechanics. Then we can speak
about the loss of stability of the original bush $B[G_D]$ and its
transformation into a larger bush $\tilde{B}[\widetilde{G_D}]$
with $\widetilde{G_D} \subset G_D$.

It was found from numerical experiments that the boundaries of
stable (unstable) domains for bushes of vibrational modes depend
not only on the initial energy $E_0$, but on the {\it way}
of excitation, as well. Because of this reason, we will {\it fix}
the way of initial excitation.\footnote{Similar results can be
also obtained for other ways of excitation.}

Let us first discuss the loss of stability of the bushes from
Table~\ref{2a}. The root mode of each of these bushes is
determined by a {\it single} time-dependent coefficient
$\mu_i^{(j)}(t)$ whose initial value at $t=t_0$ we will denote by
the symbol $\mu_0$ $(\mu_0 \equiv \mu_i^{(j)}(t_0))$. At the
initial instant $t=t_0$, we fix the coordinates of all particles of
the mechanical system in such a way that their {\it displacements}
correspond to the appropriate root mode with amplitude $\mu_0$,
while their velocities are equal to {\it zero}. Namely this choice
of initial conditions determines the above mentioned {\it way} of
excitation of a given bush.

Using these initial conditions we solve numerically the exact
dynamical equations of the considered mechanical system with 18
degrees of freedom, and analyze the set of nonzero modes
$\mu_i^{(j)}(t)$ in the decomposition (\ref{eq2}) of the vector
${\bf X}(t)$ obtained as a result of this solving. Then we
gradually increase the value $\mu_0$ and repeat the procedure just
described until the number of nonzero modes  $\mu_i^{(j)}(t)$, at
some value $\mu_0=R$, becomes larger than that of the bush
$B[G_D]$. We will refer to $R$ as the {\it threshold} of stability
of the given bush. Obviously, in such a way we obtain the upper
boundary of the first stability region of $B[G_D]$ in the
one-dimensional space of all possible values $\mu_0 \; (0< \mu_0
\leq R)$.\footnote{Note that we do not study the {\it other}
possible regions of stability of the given bush $B[G_D]$ in the
present paper.}

The threshold values $R$, with the appropriate numerical
accuracy,\footnote{We hope that numerical errors do not exceed
unity in the last digits of numbers in our tables.} for different
bushes for the octahedral structure without a centered particle
can be found in the last column of Table~\ref{2a}. For example,
the threshold $R=0.028$ corresponds to bush $B3[D'_{2d}]$. It
means that below this threshold only three nonzero modes can be
revealed in the decomposition of the appropriate vector \({\bf
X}(t) : \mu_3^{(8)}(t)\) (root mode) and
\(\mu_1^{(1)}(t),\mu_1^{(5)}(t)\) (secondary modes), while beyond
this value of $R$ some new modes $\mu_i^{(j)}(t)$ appear in the
decomposition (\ref{eq2}) of the vector \({\bf X}(t)\). We treat
this fact as evidence of the loss of stability of bush
$B3[D'_{2d}]$ and its transformation to a new bush of higher
dimension. Note that in the present paper, we do not determine
what this new bush is, because realization of such a goal requires
a sufficient time due to some computational errors problems.

We already discussed the bush $B6[C'_{2v}]$ from Table~\ref{2a}
which can be excited either by an initial excitation of the root mode
$\mbox{\boldmath$\varphi$}_1^{(8)}+
\mbox{\boldmath$\varphi$}_2^{(8)}$ of the irrep $\Gamma_8$ with
the amplitude $\mu_1^{(8)}(t)$, or by excitation of the root mode
$\mbox{\boldmath$\varphi$}_1^{(10)}-
\mbox{\boldmath$\varphi$}_2^{(10)}$ of the irrep $\Gamma_{10}$
with the amplitude $\mu_1^{(10)}(t)$. In such cases, we find {\it
two} different threshold values $R_1$ and $R_2$ corresponding
to the two different root modes.

A more complicated computational problem arises when we analyze
the domains of stability for some bushes of modes from
Table~\ref{2b}. Indeed, the root modes of bushes $B10[D_{2h}]$,
$B11[D'_2]$, $B12[C_{2v}]$, $B13[C''_{2v}]$, $B14[C'_{2}]$,
$B16[C'_{s}]$, $B17[C_3]$, include {\it two} independent functions
$\mu_i^{(j)}(t)$. For some of the above bushes these functions are
associated with basis vectors of one and the same irrep; in other
cases they are associated with basis vectors of two different
irreps. Therefore, we must now study the boundary of the stability
domain in a two-dimensional space of initial values, say,
$\mu_1(t_0)$ and $\mu_2(t_0)$, of the two appropriate
time-dependent variables $\mu_i^{(j)}(t)$. Since studying of the
form of such a boundary, by means of numerical experiments,
requires a lot of computational time, we restrict ourselves to the
determination of only the radius $R'$ of a circle which with
assurance lies in the stability domain. Namely these radii we give
in the last column of Table~\ref{2b} as certain estimation
characteristics of stability domains of the appropriate bushes of
vibrational modes.

In Table~\ref{2b}, as well as in Table~\ref{2a}, we
encounter the case where a given bush can be excited by
using different choices of the root mode, but now each of these
modes is determined by two amplitudes $\mu_i^{(j)}(t)$. This is
the bush $B15[C_s]$ whose root modes are $\mu_1^{(8)}(t)
\mbox{\boldmath$\varphi$}_1^{(8)}+ \mu_2^{(8)}(t)
\mbox{\boldmath$\varphi$}_2^{(8)}$ or $\mu_1^{(10)}(t)
\mbox{\boldmath$\varphi$}_1^{(10)}+ \mu_2^{(10)}(t)
\mbox{\boldmath$\varphi$}_2^{(10)}$. Because of this reason, two
different radii $R'_1$ and $R'_2$ are given in Table~\ref{2b}
corresponded to the above two variants of the root mode.

Finally, the root mode of the last bush $B18[C_i]$ in
Table~\ref{2b} depends on three functions $\mu_1^{(7)}(t),
\mu_2^{(7)}(t), \mu_3^{(7)}(t)$ and, therefore, we must study, in
this case, the boundary of the stability domain in
three-dimensional space. We give for this bush the radius $R''$ of
a sphere which with certainty belongs to the stability domain.

\subsection{Numerical results on bush stability for octahedral
structure without centered particle}

Up to this point our consideration was based on
group-theoretical methods which are independent of the type
of interactions between particles in the mechanical system. Now we
intend to discuss the findings on bush stability
using for this purpose the concrete type of interactions described
by the Lennard-Jones potential already discussed in
Sec.~\ref{exact} (see Eq.~(\ref{eq10})).

First of all, we must examine the stability of the octahedral
structure depicted in Fig.1 in the equilibrium state. To this end, we
found squares of all eigenfrequencies in the harmonic
approximation with the aid of the MAPLE package and made their
classification in accordance with the irreps of the group $O_h$.
We obtained the following results:
\[\omega^2[\Gamma_1]=61.082;\qquad \omega^2[\Gamma_5]=13.835;\]
\begin{equation}
\label{100} \omega^2[\Gamma_7]=31.498; \qquad
\omega^2[\Gamma_8]=15.346;
\end{equation}
\[\omega^2[\Gamma_9]=0; \qquad \omega^2[\Gamma_{10}]=46.843;0.\]
Here $\omega^2[\Gamma_j]$ is a square of the frequency
corresponding to the irrep $\Gamma_j$.

The irreps $\Gamma_2,\Gamma_3,\Gamma_4$ and $\Gamma_6$ are not
contained in the decomposition of the full vibrational
representation $\Gamma_{\rm mech}$ of the octahedral structure
and, as a consequence, there are no vibrational frequencies
corresponding to them.

Note, $\omega^2[\Gamma_9]$ and the last value of
$\omega^2[\Gamma_{10}]$ are equal to zero. They correspond,
respectively, to pure rotation and pure translation of the
mechanical system as a whole, and must be excluded from the
further discussion because we consider vibrational regimes
only.\footnote{Remember that movements of rotation-vibration type
were excluded from our consideration in the present paper.} All
other $\omega^2[\Gamma_j]$ are {\it positive} and this proves that
the equilibrium state of our mechanical system is stable.

Let us also write down the value $a_0$ of the edge of the regular
octahedron depicted in Fig.\ref{fig.1} in equilibrium:
\begin{equation}
\label{101} a_0=1.117.
\end{equation}
Hereafter, we use such dimensionless units that $A=1$ and $B=1$ in
the formula (\ref{eq10}) for the Lennard-Jones potential (see
Sec.~\ref{exact}).

Now we consider our results on the stability of dynamical regimes
described by bushes of vibrational modes. The computational
scheme, outlined in the previous section, and fourth order
Runge-Kutta method were used for appropriate numerical
experiments. The threshold values $R,R',R''$ characterizing the
size of stability domains for all bushes are given in the last
column of Tables~\ref{2a},~\ref{2b}. It can be seen from these
tables that the size of stability domains differs very
considerably for different bushes: $0.001\leq R \leq 0.078, 0.001
\leq R' <0.027$.

The one-dimensional bush $B1[O_h]$, consisting of the breathing
mode only, possesses the minimal ``stability reserve": $R=0.001$.
Obviously, this is a very small value for $R$, because the
displacements of particles corresponding to it are
approximately equal to 0.0005, while the equilibrium edge $a_0$ of
our octahedron is equal to 1.117 according to Eq.(\ref{101}).
Thus, the bush $B1[O_h]$ exhibits very weak stability for
the octahedral structure without a centered particle.
Nevertheless, if a particle in the center of the octahedron
depicted in Fig.\ref{fig.1} is present, the threshold
value $R$ for this bush can be up to 100--1000 times greater,
depending on the force properties of the centered particle
(see Sec.~\ref{seq5.4}).

The maximal value $R$ corresponds to the bush $B9[D_{3d}] \;
(R=0.078)$, demonstrating how strong can be the difference between
the stability properties of dynamical regimes with different character
of displacements of individual particles (note that using
Tables~\ref{2a},~\ref{2b},~\ref{1}, we can imagine the character
of movement of the mechanical system for any given bush).

Our main result can be formulated as follows: {\it all} bushes of
vibrational modes possess stability domains of a {\it finite
size} for the octahedral structure with the Lennard-Jones potential.

\subsection{Interactions between different modes}
\label{mu-space}

In many physical papers, pair interactions between modes are
discussed (compare the pho\-non-phonon interactions in the
presence of anharmonic terms in the appropriate Hamiltonian). In a
rigorous sense, such a setting of the problem is incorrect and can
be used only as a first step of investigation, for example, to
obtain some estimations. Indeed, we can formally write the
Lagrange equations describing dynamics of only two arbitrarily
chosen modes, but this new two-dimensional dynamical system may
not correspond to any real dynamical regime in the original
mechanical system.

The importance of the bush theory is brought about, above all, by the
following consequence: choosing a given bush, we can
guarantee that the full collection of its modes is {\it closed},
i.e. that the appropriate dynamical regime including {\it only
these} modes {\it can really exist} in the considered mechanical
system. As a consequence, the Lagrange equations, obtained for
the complete set of the modes of a given bush, are exact.

Nevertheless, studying the interactions between
pairs of different normal modes may be useful in certain cases as
an approximate method for revealing some essential properties of
nonlinear dynamics of the mechanical system. Let us illustrate
this idea using the following examples.

1. We consider a possibility of the parametric excitation of
``sleeping" modes\footnote{We mean by this term the modes which do
not belong to the bush generated by a given root mode
($\mu_1^{(1)}(t)$, in our case) and which were not excited at the
initial instant $t=t_0$.}, belonging to the irreps
$\Gamma_5,\Gamma_7,\Gamma_8$ and $\Gamma_9$, brought about by
their {\it pair interactions} with the breathing mode
$\mu_1^{(1)}(t)$. In other words, we must find the threshold
values $\mu_1^{(1)}(0) \equiv \mu_c$ beyond which each of these
previously sleeping modes can appear because of its interaction
with the ``active" mode $\mu_1^{(1)}(t) \neq 0$. The following
results were obtained by means of numerical experiments:
\[\Gamma_5[\mu_1^{(5)}(t),\mu_2^{(5)}(t)- \mu_c = 0.020],\]
\begin{equation}
\label{102} \Gamma_7[\mu_1^{(7)}(t),\mu_2^{(7)}(t),\mu_3^{(7)}(t)-
\mu_c = 0.173],
\end{equation}
\[\Gamma_8[\mu_1^{(8)}(t),\mu_2^{(8)}(t),\mu_3^{(8)}(t)- \mu_c = 0.001],\]
\[\Gamma_{10}[\mu_1^{(10)}(t),\mu_2^{(10)}(t),\mu_3^{(10)}(t)- \mu_c = 0.333].\]
Note that modes listed in square brackets next to the symbol of
the appropriate irrep possess the identical threshold value
$\mu_c$.

According to Eq.(\ref{102}), the minimal threshold value $\mu_c$
corresponds to the modes of the irrep $\Gamma_8$ $(\mu_c=0.001)$,
and this value coincides with the threshold of stability
$R=0.001$ of the one-dimensional bush $B1[O_h]$ pointed out in
Table~\ref{2a}. It is essential, that when we find this value $R$
for $B1[O_h]$ by means of the numerical procedure described in
Sec.\ref{set_prob}, we cannot reveal the {\it cause} of such a small
value of $R$. Now we can assert that the bush $B1[O_h]$ loses its
stability  because of the interactions with the modes namely of
the irrep $\Gamma_8$ and, therefore, this bush must transform into
a certain bush, containing these modes, such as $B3[D'_{2d}],
B6[C'_{2v}]$, etc., but not into such bushes as $B2[D_{4h}],
B4[C_{2v}]$, etc.

2. Let us discuss the threshold value $R=0.009$ of the stability
of the bush $B2[D_{4h}]$ (see Table~\ref{2a}) in terms of pair
interactions between different modes. This bush contains two
modes: the root mode $\mu_1^{(5)}(t)$ and the secondary mode
$\mu_1^{(1)}(t)$. We can find the threshold of the initial value
$\mu_1^{(5)}(0)\equiv \mu_c$ beyond which the active mode
$\mu_1^{(5)}(t)$ excites, by means of its parametric action,
other modes which were previously sleeping.  (There is one and only one mode
--- $\mu_1^{(1)}(t)$ --- which is brought into the
vibrational process by {\it force} originating from the active mode
$\mu_1^{(5)}(t)$.) We obtained the following results:
\begin{equation}
\label{103} \Gamma_5[\mu_2^{(5)}(t) - \mu_c = 0.273],
\end{equation}
\[\Gamma_{7}[\mu_1^{(7)}(t),\mu_2^{(7)}(t)-\mu_c=0.560;\mu_3^{(7)}(t)- \mu_c = 0.403],\]
\[\Gamma_8[\mu_1^{(8)}(t)-\mu_c=0.344;\mu_2^{(8)}(t)-\mu_c=0.344;\mu_3^{(8)}(t)- \mu_c = 0.455],\]
\[\Gamma_{10}[\mu_1^{(10)}(t)-\mu_c=0.616; \mu_2^{(10)}(t)-\mu_c=0.616; \mu_3^{(10)}(t)- \mu_c = 0.577].\]

Note that unlike the case where the breathing mode was active (see
Eq.(\ref{102})), threshold values $\mu_c$ for the excitation of
sleeping modes by the active mode $\mu_1^{(5)}(t)$ can be {\it
different} for the different modes of one and the same
multidimensional irrep.

A remarkable fact can be revealed in
analyzing the above threshold values given in
Eq.(\ref{103}). Indeed, all $\mu_c$,
corresponding to the initial value of the root mode
$\mu_1^{(5)}(t)$, are {\it essentially greater} than the threshold
$R=0.009$ for the loss of stability of the considered bush
$B2[D_{4h}]$. On the other hand, as we already know, the threshold
for the parametric excitation of the modes associated with the
irrep $\Gamma_8$ by action from the active mode $\mu_1^{(1)}(t)$
is a very small value: $\mu_c\equiv\mu_1^{(1)}(0)=0.001$. This
fact suggests that the sufficiently weak stability of
the bush $B2[D_{4h}]$ brought about by the parametric excitation
of some sleeping modes originated not from the root, but from the
{\it secondary} mode of our bush! Indeed, if we consider dynamics
of the bush $B2[D_{4h}]$ when $\mu_1^{(5)}(0)$ are slightly higher
than 0.009, the value of its secondary mode
$\mu_1^{(1)}(t)$ does reach the threshold value 0.001 which can
lead to excitation of the modes of the irreps $\Gamma_8$.

Thus, we encounter now the case where a given bush loses its
stability because of the phenomenon similar to the parametric
resonance induced by its {\it secondary} mode, and this property
demonstrates, in particular, that a bush is an indivisible
dynamical object.

\subsection{Stability of bushes of vibrational modes for
octahedral structure with the centered particle}
\label{seq5.4}

First of all, let us note that all octahedral molecules, known to
us at the present time, possess an atom in the center of the
octahedron (see Fig.\ref{fig.1}). This fact suggests that the
stability of such structures can be greater than
that of the mechanical system considered in the previous sections.
Taking into account this hint, we examined bush stability for the
mechanical structure depicted in Fig.\ref{fig.1}, supposing that
the particle in the center of the octahedron is described by the
Lennard-Jones potential {\it different} from that of six
peripheral particles.

Let us assume that the Lennard-Jones potential of the form
\begin{equation}
\widetilde{U}(r)=\frac{1}{r^{12}}-\frac{B}{r^6} \label{150}
\end{equation}
with $B>1$ corresponds to the centered particle, while all other
particles are described by the potential (\ref{150}) with $B=1$.
Such an assumption provides us a possibility to make the attractive part
of the potential of the centered particle greater than that of
peripheral atoms in spite of the same repulsive part.

We will consider only several bushes of modes for the centered
structure to show that the size of the stability domain can be
increased essentially by means of the appropriate choice of the
value B. Note for the beginning of the discussion, that all bushes
from Tables~\ref{2a},\ref{2b} are relevant bushes for centered
structure, as well as for that without the particle in the center
of the octahedron, if this particle is
fixed.\footnote{If the centered particle can move, we obtain some
{\it additional} bushes to those in Tables~\ref{2a},~\ref{2b}.}

Let us discuss some results of the appropriate numerical
experiments. The boundary of the stability domain for the bush
$B1[O_h]$ can be increased from the threshold value $R=0.001$ (see
Table~\ref{2a}) up to the value $R=1.01$ for the case of the
centered structure with $B=5.5$. (It is interesting that the
function $R(B)$ for this bush is not monotonic). The threshold
values $R$ for other considered bushes also become larger for
$B=5.5$ in comparison with those for $B=1$. For example, we
obtained $R=0.011$ for bush $B2[D_{4h}]$ instead of $R=0.009$ and
$R=0.118$ for bush $B4[C_{4v}]$ instead of $R=0.003$. Note that
the dependence of the boundary $R$ of stability domains on the
value $B$ is essentially different for different bushes.

We can conclude that the stability of bushes of vibrational modes
for the centered structure can be increased, and for some bushes
substantially increased, in comparison to that of the mechanical system
without the centered particle on account of the appropriate choice of
the constant $B$ in the Lennard-Jones potential (\ref{150}).

\section{Summary}

The specific dynamical objects --- {\it bushes of normal modes} ---
whose theory was developed in \cite{Dan1,Dan2,PhysD}, are
discussed in the present paper for octahedral mechanical systems
with point masses. Being a certain collection of modes, a given
bush possesses some symmetry group which is a subgroup of the
symmetry group of the mechanical system in equilibrium. The bush
can be excited under definite initial conditions, and the energy
of the initial excitation turns out to be trapped in this bush.

We found that there exist 18 different by symmetry bushes of
vibrational modes for the considered mechanical system. We
examined the stability of all these bushes for the case where
interactions between particles are described by the Lennard-Jones
potential. The main results are listed in Tables \ref{2a} and
\ref{2b}.

It was proved that all of the above mentioned bushes possess stability
domains of {\it finite size} and, therefore, they really
can be treated as certain physical objects.

\section*{Appendix 1. The point group $O_h$ and its irreducible representations}

The group $O_h$ consists of 48 symmetry elements. We can obtain
all these elements with the aid of the different products of a
certain set of generators. The minimal number of generators for
group $O_h$ is equal 2, but it is more convenient, for our purpose,
to use an extended set of generators. In notation of
the handbook by Kovalev~\cite{Kovalev}, we chose the following
five generators: $h_2, h_3, h_5, h_{13}, h_{25}$. Here  $h_2, h_3$
and $h_{13}$ are $180^\circ$ rotations about the two-fold axes
[100], [010] and [$\bar 1$10], respectively; $h_5$ is a $240^\circ$
rotation about the three-fold axis [111], and $h_{25}$ is
the inversion. Using these generators we can obtain, step by step, the
following chain of groups: $C_2[h_2]\to D_2[h_2, h_3]\to
T[h_2, h_3, h_5]\to O[h_2, h_3, h_5, h_{13}]\to O_h$. Each
group is obtained from the preceding group by adding one new
generator.\footnote{We use Schoenflies symbols for the symmetry
groups and point out their generators in square brackets next to
these symbols.}

The group $O_h$ has ten irreducible representations
$\Gamma_j\; (j=1,\ldots,10)$. Matrices of the generators of this
group for all irreps are given below in Table~\ref{4}.
\begin{table}[htb]
\caption{Matrices of the generators of the group $O_h$} \label{4}
$$
\begin{array}{|c|c|c|c|c|c|}
\hline
 & h_2 & h_3 & h_5&h_{13}&h_{25}\\ \hline
 \Gamma_1 & 1&1&1&1&1\\
 \Gamma_2 &1&1&1&1&-1\\
 \Gamma_3 &1&1&1&-1&1\\
 \Gamma_4 &1&1&1&-1&-1\\
 \Gamma_5 &\left( \begin{array}{cc} 1&0 \\ 0&1 \end{array} \right)&
 \left( \begin{array}{rr} 1&0 \\ 0&1 \end{array} \right)&
 \left( \begin{array}{rr} -\frac{1}{2}&\frac{\sqrt{3}}{2} \\ -\frac{\sqrt{3}}{2}&-\frac{1}{2} \end{array} \right)&
 \left( \begin{array}{rr} 1&0 \\ 0&-1 \end{array} \right)&
 \left( \begin{array}{rr} 1&0 \\ 0&1 \end{array} \right)\\
 \Gamma_6 &\left( \begin{array}{rr} 1&0 \\ 0&1 \end{array} \right)&
 \left( \begin{array}{rr} 1&0 \\ 0&1 \end{array} \right)&
 \left( \begin{array}{rr} -\frac{1}{2}&\frac{\sqrt{3}}{2} \\ -\frac{\sqrt{3}}{2}&-\frac{1}{2} \end{array} \right)&
 \left( \begin{array}{rr} 1&0 \\ 0&-1 \end{array} \right)&
 \left( \begin{array}{rr} -1&0 \\ 0&-1 \end{array} \right)\\
 \Gamma_7 &\left( \begin{array}{rrr} 1&0&0 \\ 0&-1&0 \\ 0&0&-1 \end{array} \right)&
 \left( \begin{array}{rrr} -1&0&0 \\ 0&1&0 \\ 0&0&-1 \end{array} \right)&
 \left( \begin{array}{rrr} 0&1&0 \\ 0&0&1 \\ 1&0&0 \end{array} \right)&
 \left( \begin{array}{rrr} 0&1&0 \\ 1&0&0 \\ 0&0&1 \end{array} \right)&
 \left( \begin{array}{rrr} 1&0&0 \\ 0&1&0 \\ 0&0&1 \end{array} \right) \\
 \Gamma_8 &\left( \begin{array}{rrr} 1&0&0 \\ 0&-1&0 \\ 0&0&-1 \end{array} \right)&
 \left( \begin{array}{rrr} -1&0&0 \\ 0&1&0 \\ 0&0&-1 \end{array} \right)&
 \left( \begin{array}{rrr} 0&1&0 \\ 0&0&1 \\ 1&0&0 \end{array} \right)&
 \left( \begin{array}{rrr} 0&1&0 \\ 1&0&0 \\ 0&0&1 \end{array} \right)&
 \left( \begin{array}{rrr} -1&0&0 \\ 0&-1&0 \\ 0&0&-1 \end{array} \right) \\
 \Gamma_9 &\left( \begin{array}{rrr} 1&0&0 \\ 0&-1&0 \\ 0&0&-1 \end{array} \right)&
 \left( \begin{array}{rrr} -1&0&0 \\ 0&1&0 \\ 0&0&-1 \end{array} \right)&
 \left( \begin{array}{rrr} 0&1&0 \\ 0&0&1 \\ 1&0&0 \end{array} \right)&
 \left( \begin{array}{rrr} 0&-1&0 \\ -1&0&0 \\ 0&0&-1 \end{array} \right)&
 \left( \begin{array}{rrr} 1&0&0 \\ 0&1&0 \\ 0&0&1 \end{array} \right) \\
 \Gamma_{10} &\left( \begin{array}{rrr} 1&0&0 \\ 0&-1&0 \\ 0&0&-1 \end{array} \right)&
 \left( \begin{array}{rrr} -1&0&0 \\ 0&1&0 \\ 0&0&-1 \end{array} \right)&
 \left( \begin{array}{rrr} 0&1&0 \\ 0&0&1 \\ 1&0&0 \end{array} \right)&
 \left( \begin{array}{rrr} 0&-1&0 \\ -1&0&0 \\ 0&0&-1 \end{array} \right)&
 \left( \begin{array}{rrr} -1&0&0 \\ 0&-1&0 \\ 0&0&-1 \end{array} \right) \\ \hline
\end{array}
$$
\end{table}

\section*{Acknowledgements}

The authors thank Prof. V.P. Sakhnenko for useful discussions and
Prof. H.T. Stokes for his great help in correction of the text of
the present paper.

\begin{figure}[p]
\centering \unitlength=0.12pt
\begin{picture}(1472,1068)
\end{picture}
\caption{Octahedral mechanical system.} \label{fig.1}
\end{figure}

\end{document}